\documentclass[11pt,a4paper]{article}
\pdfoutput=1

\usepackage{jcappub}
\usepackage{amsmath,amssymb,color,bm, graphicx}
\usepackage{enumerate,xstring, hyperref}

\def\be{\begin{equation}}
\def\ee{\end{equation}}
\def\bea{\begin{eqnarray}}
\def\eea{\end{eqnarray}}
\newcommand{\vs}{\nonumber\\}
\def\ba#1\ea{\begin{align}#1\end{align}}
\def\bg#1\eg{\begin{gather}#1\end{gather}}

\def\iMpch{\,h\,{\rm Mpc}^{-1}}
\newcommand{\fkone}{k\frac{P^\prime_L(k)}{P_L(k)}}
\newcommand{\fktwo}{k^2\frac{P^{\prime\prime}_L(k)}{P_L(k)}}
\newcommand{\fkonenl}{k\frac{P^\prime_m(k)}{P_m(k)}}
\newcommand{\fktwonl}{k^2\frac{P^{\prime\prime}_m(k)}{P_m(k)}}

\def\knl{k_\text{NL}}
\def\Ote{\hat{\varPi}}

\newcommand{\s}{\sigma}
\newcommand{\refeq}[1]{Eq.~(\ref{eq:#1})}

\newcommand{\reffig}[1]{Fig.~\ref{fig:#1}}          
          
\newcommand{\refsec}[1]{Sec.~\ref{sec:#1}}

\def\Plin{P_{\rm L}}

\renewcommand{\v}[1]{\bm{#1}}

\newcommand{\vx}{\v{x}}

\newcommand{\vk}{\v{k}}

\newcommand{\vp}{\v{p}}

\newcommand{\<}{\langle}
\renewcommand{\>}{\rangle}

\renewcommand{\d}{\delta}

\def\be{\begin{equation}}
\def\ee{\end{equation}}
\def\ben{\begin{eqnarray}}
\def\een{\end{eqnarray}}
\def\ba{\begin{array}}
\def\ea{\end{array}}

\def\ba#1\ea{\begin{align}#1\end{align}}

\newcommand{\bq}{\begin{eqnarray}}
\newcommand{\eq}{\end{eqnarray}}
\newcommand{\bes}{\begin{subequations}}
\newcommand{\ees}{\end{subequations}}

\def\R{\mathcal{R}}

\def\P{\mathcal{P}}

\def\O{\mathcal{O}}

\def\A{\mathcal{A}}
\def\B{\mathcal{B}}
\def\C{\mathcal{C}}
\def\fii{F_2}
\def\fiii{F_3}
\def\kunit{\:h\,{\rm Mpc}^{-1}}

\def\oneloop{1-{\rm loop}}

\makeatletter
\newlength{\apb@width}
\newcommand{\autoparbox}[2][c]{\settowidth{\apb@width}{#2}\parbox[#1]{\apb@width}{#2}}
\newcommand{\includegraphicsbox}[2][]{\autoparbox{\includegraphics[#1]{#2}}}
\makeatother

\DeclareMathOperator{\cov}{Cov}

\newcommand{\comment}[1]{}

\def\Mpch{\,h^{-1}\,{\rm Mpc}}

\begin{document}

\title{Response Approach to the Matter Power Spectrum Covariance}

\author{Alexandre Barreira and}
\emailAdd{barreira@MPA-Garching.MPG.DE}

\author{Fabian Schmidt}
\emailAdd{fabians@MPA-Garching.MPG.DE}

\affiliation{Max-Planck-Institut f{\"u}r Astrophysik, Karl-Schwarzschild-Str. 1, 85741 Garching, Germany}

\abstract{We present a calculation of the matter power spectrum covariance matrix $\cov(\vk_1,\vk_2)$ that uses power spectrum responses to accurately describe the coupling between large- and small-scale modes beyond the perturbative regime. These response functions can be measured with (small-volume) N-body simulations, which is why the response contributions to the covariance remain valid and predictive at all orders in perturbation theory. A novel and key step presented here is the use of responses to compute loop contributions with soft loop momenta, which extends the application of the response approach beyond that of previously considered squeezed $n$-point functions. The calculation presented here does not involve any fitting parameters. When including response-type terms up to 1-loop order in perturbation theory, we find that our calculation captures the bulk of the total covariance as estimated from simulations up to values of $k_1, k_2 \sim 1\kunit$.  Moreover, the prediction is guaranteed to be accurate whenever the softer mode is sufficiently linear, $\min\{k_1,k_2\} \lesssim 0.08\kunit$.  We identify and discuss straightforward improvements in the context of the response approach, which are expected to further increase the accuracy of the calculation presented here.}


\date{\today}

\maketitle
\flushbottom

\section{Introduction}\label{sec:intro}

The unprecedented statistical precision that upcoming large-scale structure surveys are expected to attain requires cosmologists to develop equally precise methods to predict the various observables. The simplest and most widely applied way to describe the statistical information encoded in the large-scale structure is via the $2$-point correlation function \cite{peebles:1980}. This includes the $2$-point galaxy correlation function, or $2$-point correlations of cosmic shear maps in the case of gravitational lensing, as well as their cross-correlation. The starting point to predicting both these observables is the $2$-point correlation function of matter $\xi_m$, or its Fourier transform, the power spectrum $P_m$. The matter power spectrum is very well understood in the context of gravity-only N-body simulations (that is, neglecting baryonic effects on the total matter distribution). The simulation requirements for a given pre-specified precision have been studied in Ref.~\cite{2016JCAP...04..047S}, and simulations have also allowed for the calibration of semi-analytical models such as {\sc Halofit} \cite{2012ApJ...761..152T} and construction of efficient interpolations such as that of the {\sc Coyote} project \cite{emulator}. Baryonic effects are known to have an impact on the small-scale matter power spectrum \cite{rudd/zentner/kravtsov,2014Natur.509..177V, 2016MNRAS.461L..11H} and work on modeling these effects has also been carried out recently \cite{zentner/rudd/hu, mohammed/seljak, 2015MNRAS.454.1958M, 2016PhRvD..94f3508S}.

An accurate model of the matter power spectrum alone is however insufficient to properly exploit upcoming surveys, especially when inferring cosmological parameter values from the data, for which one also needs accurate determinations of the covariance matrix of the power spectrum, 
\bq\label{eq:covdefintro}
\cov({\vk_1, \vk_2}) \equiv  \big< \hat{P}_m(\vk_1) \hat{P}_m(\vk_2) \big> - \big< \hat{P}_m(\vk_1) \big>\big< \hat{P}_m(\vk_2) \big>,
\eq
in order to quantify the statistical error of the measurements. In the equation above, angle brackets denote ensemble averages and $\hat{P}_m(\vk)$ is an estimate of the matter power spectrum within some wavemode bin centered at $\vk$. The power spectrum covariance, hereafter simply referred to as \emph{matter covariance}, therefore measures the correlation between the power at wavemodes $\vk_1$ and $\vk_2$. For Gaussian initial conditions, different Fourier modes evolve independently in the linear stages of structure formation. In this regime, only the diagonal terms are non-vanishing and they are trivially related to the matter power spectrum itself. At later stages, nonlinear structure formation effectively couples different Fourier modes, which leads to important off-diagonal ($k_1\neq k_2$) terms in $\cov({\vk_1, \vk_2})$ through a special configuration of the matter trispectrum (the Fourier transform of the $4$-point correlation function), which we will describe in more detail below. The matter trispectrum is a cumbersome quantity to predict, which is why our current knowledge of the covariance matrix is far poorer than that of the power spectrum. Given that inaccurate estimates of the covariance matrix can result in wrong interpretations of the data (see e.g.~Refs.~\cite{2007MNRAS.375L...6S, 2011MNRAS.416.1045K, 2013MNRAS.432.1928T, 2013PhRvD..88f3537D, 2016MNRAS.458.4462B}), this naturally motivates research in obtaining accurate theoretical predictions of the covariance, including its dependence on cosmological parameters \cite{2009A&A...502..721E, 2012ApJ...760...97L, 2013JCAP...11..009M, 2015JCAP...12..058W}, as well as on baryonic effects.  This is important for current data sets, but even more so for future large-volume surveys such as Euclid \cite{2011arXiv1110.3193L}, LSST \cite{2012arXiv1211.0310L} and DESI \cite{2013arXiv1308.0847L}.

One frequently employed tool to estimate $\cov({\vk_1, \vk_2})$ are direct estimates of the covariance via Eq.~(\ref{eq:covdefintro}) using large sets of N-body simulations \cite{2006MNRAS.371.1188H, 2009ApJ...700..479T, joachimpen2011, 2011ApJ...734...76S, 2017arXiv170105690K}. This requires performing thousands of N-body simulations in order to obtain sufficient signal-to-noise in the covariance, which makes these estimates extremely costly in terms of computational resources. Estimating the covariance for many different sets of cosmological parameters therefore becomes prohibitive, as does a realistic modeling of baryonic processes.  

A complementary approach to simulation estimates is to use perturbation theory. Reference~\cite{2016PhRvD..93l3505B} presented a calculation of the trispectrum in the covariance configuration at the 1-loop level, based on the Effective Field Theory (EFT) of large-scale structure (see Ref.~\cite{porto:2016} for a review). The main limitation of such perturbative approaches is that they are only applicable to sufficiently large scales $k\lesssim\knl\approx0.3\kunit$ (at $z=0$), which limits their usefulness in the analysis of data on smaller scales.\footnote{We define the nonlinear scale $\knl$ at a given redshift $z$ through $\knl^3\Plin(\knl,z)/(2\pi^2)=1$.} There have also been attempts to develop semi-analytical phenomenological models of the covariance matrix on scales $k \ > \knl$ \cite{2011ApJ...736....8N, 2015MNRAS.453..450C, mohammed/seljak, mohammed1}, but these typically involve simplifying assumptions and/or free parameters that need to be tuned to match other covariance estimates, usually simulation-based ones, for any given cosmology considered. Moreover, systematic errors made in these phenomenological estimates are not under rigorous control, and can only be estimated through comparison with simulation-based estimates.

In this paper, our goal is to describe a calculation of the covariance matrix that combines the merits of the simulation- and perturbation theory-based approaches. More concretely, in our approach, we use perturbation theory to identify the mode-coupling terms of the non-Gaussian covariance that can be resummed with simulation-calibrated power spectrum responses (see Ref.~\cite{paper1} for an in-depth discussion and Sec.~\ref{sec:respdef} below for an overview). The power spectrum responses measure the fractional change of the local power spectrum in the presence of long-wavelength perturbations, and they can be measured accurately and non-perturbatively with only a few relatively small-volume simulations \cite{2011ApJS..194...46G, 2011JCAP...10..031B, 2016JCAP...09..007B, wagner/etal:2014, response, takada/hu:2013, li/hu/takada, lazeyras/etal, 2016arXiv161204360L, 2017arXiv170103375C} (with small computational cost compared to that of fully numerical covariance estimates). The types of mode-coupling interactions that are captured by responses are therefore those that describe the coupling between long- and short-wavelength modes. All non-response type terms are calculated using standard perturbation theory (SPT), leaving the whole calculation free from fitting parameters. Moreover, we can use higher-order perturbation theory to estimate the systematic error made in the calculation.

In Ref.~\cite{paper1}, we have presented a response-based calculation of the non-Gaussian covariance at tree level in the squeezed regime, i.e., when one of the modes is linear and sufficiently smaller than the other, which can take any other value: $k_{\rm soft} \ll k_{\rm hard}$, $k_{\rm soft} \ll \knl$, for any $k_{\rm hard}$, where
\be
k_{\rm soft} \equiv {\rm min}\{k_1, k_2\}\,, \quad k_{\rm hard} \equiv {\rm max}\{k_1, k_2\}\,.
\ee
This represents an application of the well-known relation between responses and squeezed-limit correlators (in this case, the squeezed trispectrum). In this paper, we go beyond Ref.~\cite{paper1} as we demonstrate how to use responses to resum interaction vertices that involve internal soft-loop momenta, thereby permitting an efficient and accurate evaluation of the covariance for any values of $k_1$, $k_2$, including (quite crucially) cases in which $k_1 \approx k_2$. This constitutes an example of the use of responses in the calculation of non-squeezed $n$-point functions.

After establishing some notation and summarizing the definitions of power spectrum responses and covariance in Sec.~\ref{sec:def}, the steps taken in this paper can be outlined as follows:
\begin{enumerate}
\item In Sec.~\ref{sec:NGcovtree}, working at tree level in standard perturbation theory, we show how to \emph{stitch} together the standard perturbation theory and response-based results presented first in Ref.~\cite{paper1} to fully describe the matter covariance in the regime where $k_{\rm soft} \ll \knl$ and any $k_{\rm hard}$.

\item Section \ref{sec:NGcov1loop} is devoted to the novel application of responses to calculate loop interactions involving soft loop momenta but fully nonlinear external momenta. Here, we work explicitly at the 1-loop level in perturbation theory, but also describe how to account for higher loops.

\item We compare our model results to simulation-based estimates of the angle-averaged covariance in Sec.~\ref{sec:monosims}. The level of agreement we find across the range of scales probed by the simulations suggests that the calculation presented here (which has no free parameters) captures the majority of the total matter covariance. In Sec.~\ref{sec:angles}, we also look at the prediction for the dependence of $\cov(\vk_1, \vk_2)$ on the angle between the two modes.
\end{enumerate}

The covariance calculation presented here, being based on a well-defined theoretical framework, is particularly useful as it allows us to determine exactly which contributions are being left out at a given point $(k_1, k_2)$; in particular, these are  higher-loop terms, and certain non-response-type terms. This can be used to estimate the error on the covariance prediction (the \emph{error on the error} of the matter power spectrum), as well as to guide further developments. These, as well as other concluding remarks are the subject of Sec.~\ref{sec:summ}. In particular, Fig.~\ref{fig:reg} summarizes which parts of $(k_1,k_2)$-space are already completely captured by our calculation, and which parts can benefit from further work.

In Appendix \ref{app:feynman}, we spell out the Feynman rules of cosmological perturbation theory as used in the paper. We collect the expressions to evaluate response functions, as well as the corresponding non-Gaussian covariance terms in Appendices \ref{app:ro} and \ref{app:analy}, respectively. The criterion to distinguish between squeezed and non-squeezed configurations is determined in Appendix \ref{app:fsq}. In Appendix \ref{app:derivation}, we demonstrate explicitly the equivalence between 1-loop covariance terms in standard perturbation theory and the response-based description. Finally, in Appendix \ref{app:mohammed}, we compare our covariance calculation with the prediction of the model presented in Ref.~\cite{mohammed1}.

In this paper, we assume a flat $\Lambda{\rm CDM}$ cosmology for all numerical results, with the following parameters: $h = 0.72$, $\Omega_mh^2 = 0.1334$, $\Omega_bh^2 = 0.02258$, $n_s = 0.963$, $\sigma_8(z=0) = 0.801$, $\sum m_{\nu}=0$. These are the same as those used in Ref.~\cite{blot2015} in their estimates of the covariance matrix from simulations, with which we shall compare our results with. Further, in our results below, we use the {\sc CAMB} code \cite{camb} and the {\sc Coyote} emulator \cite{emulator} to compute the linear and the nonlinear matter power spectrum, respectively.

\section{Definitions and notation}
\label{sec:def}

\subsection{Power spectrum responses}
\label{sec:respdef}

In this section, we briefly recap the definition and physical content of power spectrum responses, and display the equations that we use in the remainder of the paper. We refer the reader to Ref.~\cite{paper1} for a detailed description of the response formalism. Throughout, we only consider equal-time matter correlators, and will not write the time argument explicitly to ease the notation. The Feynman rules of cosmological perturbation theory (which shall be particularly useful in our considerations below) are summarized in Appendix \ref{app:feynman}. Further, we denote magnitudes of vectors as $k = |\vk|$ and adopt a shorthand notation for the sum of vectors: $\vk_{12\cdots n} = \vk_1 + \vk_2 + \cdots + \vk_n$. The $n$-th order matter power spectrum response $\R_n$ corresponds to the following interaction vertex
\ba
&\lim_{\{p_a\} \to 0} \left(
\raisebox{-0.0cm}{\includegraphicsbox[scale=0.8]{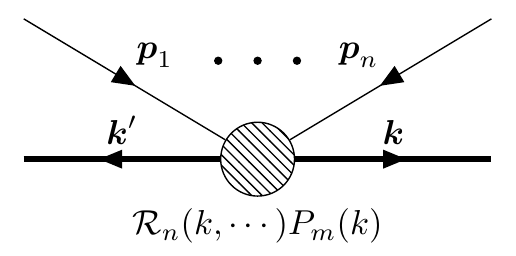}}
\right) =  \nonumber \\
\nonumber \\
& = \frac12 \R_n(k;\, \{\mu_{\vk,\vp_a}\},\, \{\mu_{\vp_a,\vp_b}\},\, \{p_a/p_b\}) P_m(k) (2\pi)^3 \d_D(\vk+\vk'- \vp_{1\cdots n})\,,
\label{eq:Rndef}
\ea
which is interpreted as the response of the nonlinear power spectrum of the small-scale (hard) mode $\vk$ to the presence of $n$ long-wavelength (soft) modes $\vp_1, ..., \vp_n$. The dashed blob is thus meant to account for the fully evolved nonlinear matter power spectrum $P_m(k)$ and all its possible interactions with the $n$ long wavelength perturbations (including loop interactions --- it is thus a \emph{resummed} vertex). In our notation, $\lim_{\{p_a\} \to 0}$ signifies that we only retain the leading contribution in the limit in which all soft momenta are taken to zero. The response $\R_n$ depends on the scale $k$, as well as on the angles between the $n$ soft modes and their angles with $\vk$. The response  also depends on the ratios of soft wavenumbers, but not on their absolute values.\footnote{These responses are to be distinguished from those measured in Refs.~\cite{neyrinck/yang, nishimichi/bernardeau/taruya}, which correspond to the derivative of the nonlinear power spectrum with respect to the initial power spectrum (i.e., not to the presence of individual large-scale perturbations).}

The diagrammatic representation of $\R_n$ helps to understand the connection of the power spectrum response with the squeezed limit of the $(n+2)$-point matter correlation function. Explicitly, attaching power spectrum propagators to the soft momentum lines in Eq.~(\ref{eq:Rndef}), we can write
\ba
\lim_{\{p_a\}\to 0} \left(
\raisebox{-0.0cm}{\includegraphicsbox[scale=0.8]{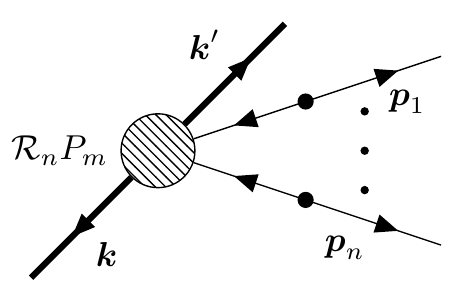}}
+ (\text{perm.}) \right)& = \<\d(\vk)\d(\vk')\d(\vp_1)\cdots \d(\vp_n)\>_{c, \R_n}
\vs
= n!\, \R_n(k;\, \{\mu_{\vk,\vp_a}\},\, \{\mu_{\vp_a,\vp_b}\},\, \{p_a/p_b\})
P_m(k) &\left[\prod_{a=1}^n\Plin(p_a)\right] \: (2\pi)^3 \d_D(\vk+\vk'+\vp_{1\cdots n})\,,
\label{eq:sqnpt}
\ea
where the $n!$ factor accounts for the permutations of the $\vp_a$. The subscript ${}_c$ denotes connected correlators, while the subscript ${}_{\R_n}$ in the $(n+2)$-connected correlator indicates that only certain contributions to the correlation function are captured by $\R_n$. There are further response-type contributions which are not included in $\<\d(\vk)\d(\vk')\d(\vp_1)\cdots \d(\vp_n)\>_{c, \R_n}$. These terms are however completely determined by lower order responses, $\R_m$, $1\leq m \leq n$, in conjunction with perturbation theory kernels. All other terms that contribute to $\<\d(\vk)\d(\vk')\d(\vp_1)\cdots \d(\vp_n)\>_c$ are small in the squeezed limit.

As described in detail in Ref.~\cite{paper1}, the $\R_n$ can be expanded in terms of all local gravitational observables associated with the $n$ long-wavelength modes, to a given order in perturbations. These observables, or operators $O$ (which can be constructed using either Lagrangian or Eulerian coordinates) form a basis $\mathcal{K}_O$ that unequivocally specifies all of the angular dependence of $\R_n$:
\be
\R_n(k;\, \{\mu_{\vk,\vp_a}\},\, \{\mu_{\vp_a,\vp_b}\},\, \{p_a/p_b\})
= \sum_O R_O(k) \mathcal{K}_O(\{\mu_{\vk,\vp_a}\},\, \{\mu_{\vp_a,\vp_b}\},\, \{p_a/p_b\})\,.
\label{eq:Rndecomp}
\ee
At any given order, there are different equivalent decompositions of the sum in \refeq{Rndecomp}, which translate into different expressions for the $\mathcal{K}_O$. For instance, Ref.~\cite{bertolini1} displays an alternative, but mathematically equivalent decomposition at $n=1$ and $n=2$. Here, we will use the Eulerian decomposition described in the main text of Ref.~\cite{paper1}. In this paper, we will only explicitly need the second-order response $\R_2 \equiv \R_2(k, \mu_1, \mu_2, \mu_{12}, p_1/p_2)$, which is a function of the hard mode $k$ (and time), the cosine angles $\mu_1 = \vp_1\cdot\vk/(p_1k)$, $\mu_2 = \vp_2\cdot\vk/(p_2k)$, $\mu_{12} = \vp_1\cdot\vp_2/(p_1p_2)$ and the ratio $p_1/p_2$. More specifically, for the application to the matter covariance, the relevant kinematic configuration corresponds to $\mu_1 = \mu$, $\mu_2 = -\mu$, $\mu_{12} = -1$ and $p_1/p_2 = 1$, in which case the expression of $\R_2$ can be given as
\bq\label{eq:PR2_anglecov}
\R_2(k; \mu, -\mu, -1, 1) &=& \left[\frac12R_2(k) + \frac23R_{K^2}(k) + \frac29 R_{K.K}(k)\right] + \left[\frac23R_{K\delta}(k) + \frac29 R_{K.K}\right]\P_2(\mu) \nonumber \\ 
&&+ \left[\frac49R_{KK}(k)\right]\left[\P_2(\mu)\right]^2 \nonumber \\
&\equiv& \A(k) + \B(k)\P_2(\mu) + \C(k)\left[\P_2(\mu)\right]^2, \nonumber \\
\eq
where $\P_\ell$ is the Legendre polynomial of order $\ell$ and the second equality serves to define the functions $\A(k)$, $\B(k)$ and $\C(k)$, which help to simplify some notation below.

The coefficients $R_O(k)$ are called \emph{response coefficients} and they correspond to the response of the local small-scale power spectrum to specific configurations of the long-wavelength perturbations. At tree level, all $R_O(k)$ can be derived by matching the definition of $\R_2$ to the squeezed four-point function, in the sense of Eq.~(\ref{eq:sqnpt}) (see Ref.~\cite{paper1} for the explicit steps of this derivation). In the nonlinear regime of structure formation, the response coefficients must be determined with the aid of N-body simulations. The first three isotropic response coefficients, $R_1$, $R_2$ and $R_3$, where $R_n(k) \equiv n! R_{\d^n}(k)$, have already been measured accurately with separate universe simulations \cite{response} (see also Refs.~\cite{2011ApJS..194...46G, 2011JCAP...10..031B, 2016JCAP...09..007B, wagner/etal:2014}). In these simulations, the presence of an exactly uniform density perturbation in the simulation volume is simulated by using the equivalence to following structure formation in a spatially curved Friedmann-Roberston-Walker spacetime \cite{CFCpaper2}. The remaining coefficients have so far not been measured in N-body simulations due to complications associated with how to model the presence of these anisotropic long-wavelength perturbations \cite{2016arXiv161001059I,2017arXiv170103375C}. In this paper, we combine the simulation measurements of the isotropic $R_O(k)$ with the nonlinear extrapolation of the anisotropic ones put forward in Ref.~\cite{paper1} (see Fig.~1 there for the numerical results). The explicit expressions for all $R_O(k)$ used in this paper are given in Appendix \ref{app:ro}.

\subsection{Matter power spectrum covariance}\label{sec:covariance}

Let $\delta(\vx)$ denote the fractional matter density contrast at $\vx$, $\delta(\vk)$ its Fourier transform (distinguished by their arguments) and $\hat{P}(\vk)$ the estimated power spectrum in a wavenumber bin centered on the Fourier mode $\vk$, in a total survey volume $V$. The matter power spectrum covariance $\cov(\vk_1, \vk_2)$ measures the correlation between the power spectrum of the modes $\vk_1$ and $\vk_2$ and is defined as (rewriting Eq.~(\ref{eq:covdefintro}))
\bq\label{eq:covdef}
\cov({\vk_1, \vk_2}) &\equiv& \cov(k_1, k_2, \mu_{12}) \equiv \big< \hat{P}_m(\vk_1) \hat{P}_m(\vk_2) \big> - \big< \hat{P}_m(\vk_1) \big>\big< \hat{P}_m(\vk_2) \big> \nonumber \\
&=& \underbrace{V^{-1}[P_m(k_1)]^2\Big[\delta_D(\vk_1+\vk_2) + \delta_D(\vk_1-\vk_2)\Big]}_{\rm Gaussian} + \underbrace{V^{-1}T_m(\vk_1, -\vk_1, \vk_2, -\vk_2)}_{\rm Non-Gaussian} \nonumber \\
&=& \cov^\text{G}(k_1, k_2, \mu_{12}) + \cov^\text{NG}(k_1, k_2, \mu_{12}),
\eq
where
\bq
\<\delta(\vk_1)\delta(\vk_2)\> &=& P_m(\vk_1) (2\pi)^3\delta_D(\vk_1+\vk_2)  \\
\<\delta(\vk_a)\delta(\vk_b)\delta(\vk_c)\delta(\vk_d)\>_c &=& T_m(\vk_a, \vk_b, \vk_c, \vk_d) (2\pi)^3 \delta_D(\vk_a + \vk_b + \vk_c+\vk_d)
\eq
define the matter power spectrum $P_m$ and trispectrum $T_m$, respectively. The latter contributes to the covariance in the so-called {\it parallelogram configuration}, $\vk_b = -\vk_a$, $\vk_d = -\vk_c$.  Note also that so far we have not restricted ourselves to the covariance of the angle-averaged power spectrum (see Eq.~(\ref{eq:averpk}) below), i.e., we allow for the covariance to depend on the angle between the two wavemodes, $\mu_{12} = \vk_1\cdot\vk_2/(k_1 k_2)$. 

As indicated in Eq.~(\ref{eq:covdef}), the two terms in the second line are broadly referred to as the Gaussian and non-Gaussian parts of the covariance, on which we comment further below. Before proceeding however, we note that, for a finite survey, Eq.~(\ref{eq:covdef}) is missing an important additional non-Gaussian contribution. This is the so-called super-sample covariance term \cite{2007NJPh....9..446T, 2009ApJ...701..945S, takada/hu:2013, li/hu/takada, 2014PhRvD..90j3530L, 2016arXiv161104723A}, which accounts for the coupling of Fourier modes inside the observed surveyed region with density fluctuations whose wavelength is larger than the typical size of the survey. Formally, this term arises from the convolution of the matter trispectrum with the survey window function. The behavior of the super-sample term is well understood and can be described using the first-order power spectrum response $\R_1$. Below, we shall compare our covariance results with estimates from standard N-body simulations, which do not include fluctuations on scales larger than the simulation box, and are therefore unable to measure the super-sample term. For this reason, we do not consider the super-sample contribution in our results, but note that its inclusion is straightforward.

For Gaussian initial conditions, the non-Gaussian contribution is only induced by nonlinear structure formation, so that the Gaussian term dominates at early times. This term correlates the power spectra of two modes only if the modes have the same magnitude and are exactly aligned, $\mu_{12}=1$ or anti-aligned $\mu_{12} = -1$. In the literature, the case of angle-averaged power spectra is that which is most commonly considered:
\bq\label{eq:averpk}
\hat{P}_m(k_1) = V_f \int_{V_s(k_1)} \frac{{\rm d}^3\vk}{V_s(k_1)}\delta(\vk)\delta(-\vk), 
\eq
where the integral is taken over a spherical shell of radius $k_1$ and width $\Delta k$, $V_s(k_1) = 4\pi k_1^2 \Delta k$, and $V_f = (2\pi)^3/V$ is the volume of a Fourier cell where $V$ is the total survey volume. 
In this case, the Gaussian part of the covariance becomes
\bq\label{eq:gauss_mono}
\cov^{\rm G}(k_i, k_j) = \frac{2}{N_k}P_m(k_i)^2\delta_{ij}\,,
\eq
where $i,j$ label bins in wavenumber, $N_k = V_s(k_i)/V_f$ is the number of Fourier modes that are averaged over in a given bin, and the Kronecker delta $\delta_{ij}$ ensures that the Gaussian term contributes only to the diagonal of the angle-averaged covariance matrix, $k_i=k_j$. Note that the Gaussian covariance depends on the size of the $k$-bins in which the spectra are measured.

The non-Gaussian part of the covariance measures the coupling between Fourier modes that is induced by nonlinear structure formation at late times. This term can also be present in the initial conditions, due to primordial non-Gaussianity, but we do not consider this case here. In the context of standard perturbation theory \cite{Bernardeau/etal:2002}, the non-Gaussian covariance $\cov^{\rm NG}$ (or the \emph{parallelogram} matter trispectrum, in the sense of Eq.~(\ref{eq:covdef})) can be expanded into its tree-, 1-loop and higher-order loop contributions,
\bq\label{eq:covngexp}
\cov^{\rm NG}(k_1, k_2, \mu_{12}) = \cov^{\rm NG}_{\rm tree}(k_1, k_2, \mu_{12}) + \cov^{\rm NG}_{\rm 1loop}(k_1, k_2, \mu_{12}) + \big({\rm higher\  loops}\big). 
\eq
This part of the covariance, which is by far the most challenging to measure and predict, is that which we wish to address specifically in this paper. The main idea behind the calculation that we perform here is that, at a given order in standard perturbation theory, we identify the mode-coupling terms that describe the interactions between hard and soft modes and resum them using power spectrum responses; all other terms can be computed as in standard perturbation theory. Up to 1-loop order, we will see how such a combination of the SPT and response approaches is capable of capturing a substantial part of the total non-Gaussian covariance.

Before proceeding with the more rigorous description of the calculation of $\cov^{\rm NG}(k_1, k_2, \mu_{12})$ in the next sections, we collect here some of the notation that is used throughout. We will use the words \emph{standard}, \emph{response} and \emph{stitched} to refer, respectively, to the SPT-based, response-based and their combined contributions to the total non-Gaussian covariance. More specifically:

\begin{enumerate}
\item \emph{Standard tree} and \emph{standard 1-loop}, which we represent as $\cov^{\rm NG}_\text{SPT-tree}$ and $\cov^{\rm NG}_\text{SPT-1loop}$, refers to the standard perturbation theory calculation (at the corresponding tree- or 1-loop levels) that does not employ any response vertices and that loses predictivity whenever any of the external momenta approach $\knl$.

\item \emph{Response tree} and \emph{response 1-loop}, which we represent as $\cov^{\rm NG}_{\R\text{-tree}}$ and $\cov^{\rm NG}_{\R\text{-1loop}}$, refers to the mode-coupling terms between hard and soft modes that exist at tree and 1-loop levels, respectively, and that can be calculated with response vertices. These contributions lose predictivity if the soft modes involved approach $\knl$, but are otherwise valid for any value of the hard modes, including in the nonlinear regime.

\item \emph{Stitched tree}, which we represent as $\cov^{\rm NG}_\text{st-tree}$, refers to a specific combination, or \emph{stitching}, of the standard and response contributions. The details of this stitching will be clarified in the sections below. Note that while we apply this procedure at tree level here, the stitching can in principle be applied at any order.
\end{enumerate}

It is also useful to organize the angular dependence of the covariance into multipoles as
\bq\label{eq:angleaver}
\cov(k_1, k_2, \mu_{12}) &=& \sum_{\ell\ {\rm even}} \cov^{\ell}(k_1, k_2) \P_\ell(\mu_{12}) \nonumber \\
\cov^{\ell}(k_1, k_2) &=& \frac{2\ell + 1}{2}\int_{-1}^1{\rm d}\mu_{12} \cov (k_1, k_2, \mu_{12}) \P_\ell(\mu_{12}).
\label{eq:covtree_exp}
\eq
The case of the monopole $\ell = 0$, $\cov^{\rm NG, \ell=0}(k_1, k_2)$, corresponds to the covariance matrix of angle-averaged spectra, which is the case we shall mostly focus on (with the exception of Sec.~\ref{sec:angles}, where we present the $\ell=2$ and $\ell=4$ predictions).

Finally, throughout we use $k_{\rm soft}$ and $k_{\rm hard}$ to denote the softest and the hardest of the two $k$-modes of the covariance, i.e., $k_{\rm soft} = {\rm min}\{k_1, k_2\}$ and $k_{\rm hard} = {\rm max}\{k_1, k_2\}$.

\section{The \emph{stitched} non-Gaussian covariance at tree level}\label{sec:NGcovtree}

In this section, we propose a calculation of the tree-level covariance that combines the standard perturbation theory result, valid only if $k_1, k_2 \ll \knl$, with the response-based description first put forward in Ref.~\cite{paper1}, which effectively extends the validity of the calculation to $k_{\rm soft} \ll \knl$, but any $k_{\rm hard}$, including in the nonlinear regime.

The SPT result for the tree-level non-Gaussian covariance \cite{1999ApJ...527....1S} is given by (see e.g.~Appendix B of Ref.~\cite{paper1} for explicit expressions for all terms of the general tree-level trispectrum)
\bq\label{eq:covtree}
V\cov_{\rm SPT-tree}^{\rm NG}(k_1, k_2, \mu_{12}) &=& 12\fiii(\vk_1, \vk_2, -\vk_2)\Plin(k_1)[\Plin(k_2)]^2 \nonumber \\
&+& 4\fii(\vk_1-\vk_2, \vk_2)^2 [\Plin(k_2)]^2\Plin(|\vk_1-\vk_2|) \nonumber \\
&+& 4\fii(\vk_1+\vk_2, -\vk_2)^2 [\Plin(k_2)]^2\Plin(|\vk_1+\vk_2|) \nonumber \\
&+& 4\fii(\vk_1-\vk_2, \vk_2)\fii(\vk_2-\vk_1, \vk_1)\Plin(k_1)\Plin(k_2)\Plin(|\vk_1-\vk_2|) \nonumber \\
&+&4 \fii(\vk_1+\vk_2, -\vk_2)\fii(\vk_1+\vk_2, -\vk_1)\Plin(k_1)\Plin(k_2)\Plin(|\vk_1+\vk_2|) \nonumber \\
&+& (\vk_1 \leftrightarrow \vk_2), \nonumber \\
\eq
where $\fii$ and $\fiii$ are the symmetrized second- and third-order standard perturbation theory kernels \cite{Bernardeau/etal:2002}. In the literature, the above equation is sometimes written in a simpler way that anticipates the angle averages that are subsequently taken, but here we opted to remain general. This standard tree level result is only expected to be a good approximation to the full non-Gaussian covariance when both $k_1$ and $k_2$ are in the linear regime, $\max\{k_1,k_2\} \ll \knl$. However, as noted already above, if $k_{\rm soft}$ is sufficiently linear and smaller than $k_{\rm hard}$, then it is possible to extend the regime of validity of the tree level calculation to nonlinear values of $k_{\rm hard}$ by making use of the response $\R_2$. This follows from noting that, by taking $n=2$, $\vp_1 = -\vp_2=\vk_{\rm soft}$ and $\vk = -\vk' = \vk_{\rm hard}$ in Eq.~(\ref{eq:sqnpt}), one obtains precisely the squeezed limit of the connected $4$-point function (or trispectrum) in the covariance configuration. The following equation provides a schematic picture of this relation:
\ba\label{eq:R2treediag}
\raisebox{-0.0cm}{\includegraphicsbox[scale=0.8]{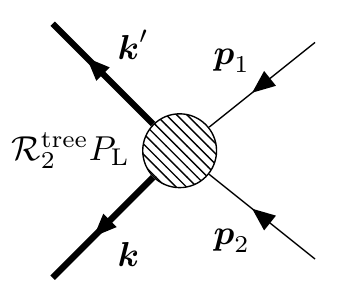}}
=\:&
\raisebox{-0.0cm}{\includegraphicsbox[scale=0.8]{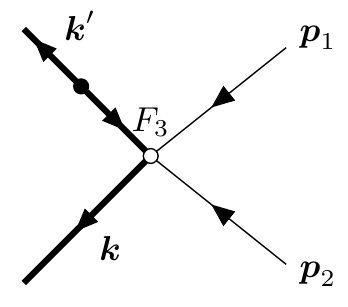}}
+
\raisebox{-0.0cm}{\includegraphicsbox[scale=0.8]{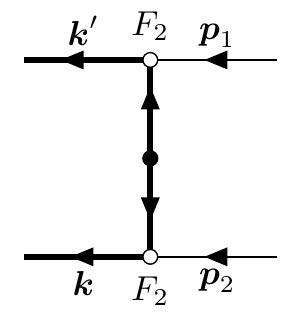}}
+ (\vk \leftrightarrow \vk')\,,
\ea
where $p_1, p_2$ are understood as much softer than $k, k'$. That is, at tree level, the $\R_2$ vertex captures the coupling described by one $F_3$ kernel in one diagram and two $F_2$ kernels in the other. The above equation is shown and used explicitly in Sec.~4.2 of Ref.~\cite{paper1} to derive the shape of $\R_2$ at tree level. By replacing the tree-level $\R_2$ with its simulation-calibrated shape, then one effectively extends (or resums) the interactions on the right-hand side of Eq.~(\ref{eq:R2treediag}) to all orders in perturbation theory in the hard mode. Referring the reader to Ref.~\cite{paper1} for more details, here we limit ourselves to showing the final response-based result, which is given by
\bq\label{eq:resptreecov}
\cov^{\rm NG}_{\R\text{-tree}}(k_1, k_2, \mu_{12}) &=& V^{-1}\,2 \R_2(k_{\rm hard}, \mu_{12}, -\mu_{12}, -1, 1)[\Plin(k_{\rm soft})]^2P_m(k_{\rm hard}) \nonumber \\ 
&&+ \O\left(\frac{k_{\rm soft}^2}{k_{\rm hard}^2},\  \frac{k_{\rm soft}^2}{\knl^2} \right),
\eq
where the next-to-leading corrections come from non-response type interactions that are suppressed in the squeezed regime, as well as loop corrections in the soft mode that enter when $k_{\rm soft}$ is no longer much smaller than $\knl$. We can now put the above two equations together to construct our stitched tree-level covariance as follows\footnote{When writing Eqs.~(\ref{eq:covtree}), (\ref{eq:resptreecov}) and (\ref{eq:stitched}) and all following relations, we have implicitly averaged over the $k$-bins used to estimate $\hat{P}_m$. Contrary to the Gaussian case, the non-Gaussian covariance does not depend explicitly on the $k$-bin widths, and hence we omit this averaging to shorten the notation.}
\bq\label{eq:stitched}
\cov^{\rm NG}_\text{st-tree}(k_1, k_2, \mu_{12}) =
\begin{cases} 
\cov_\text{SPT-tree}^{\rm NG}(k_\text{hard}, k_\text{soft}, \mu_{12}) &,\  {\rm if}\ k_\text{soft} > f_\text{sq} k_\text{hard}\\ 
\cov_{\R\text{-tree}}^{\rm NG}(k_\text{hard}, k_\text{soft}, \mu_{12}) &, \ {\rm otherwise}
\end{cases}\;.
\eq
Thus, we use the standard tree-level expression in non-squeezed configurations, but switch to the response tree-level result in squeezed ones. The value of $f_\text{sq}$ controls the transition from the non-squeezed to the squeezed regime, and its optimal choice corresponds to a trade-off between two demands. On the one hand,  the response prediction is only accurate up to corrections of order $f_\text{sq}^2$ (\refeq{resptreecov}), and hence $f_\text{sq}$ should be chosen as small as possible. On the other hand, for $k_\text{hard}$ that approach or even exceed $\knl$, the response prediction is more accurate than the SPT-tree prediction, which makes larger values of $f_\text{sq}$ beneficial (to maximize the volume in $(k_1,k_2)$-space where the response-based result is used). In Appendix \ref{app:fsq}, we describe a procedure to determine the largest value of $f_\text{sq}$ that ensures a given accuracy of Eq.~(\ref{eq:stitched}) based on the standard tree-level covariance. From the exercise performed in Appendix \ref{app:fsq}, we take our fiducial choice to be $f_\text{sq} = 0.5$. From hereon in this paper we shall therefore dub configurations with $k_\text{soft} < k_\text{hard}/2$ as squeezed.

\begin{figure}
	\centering
	\includegraphics[width=\textwidth]{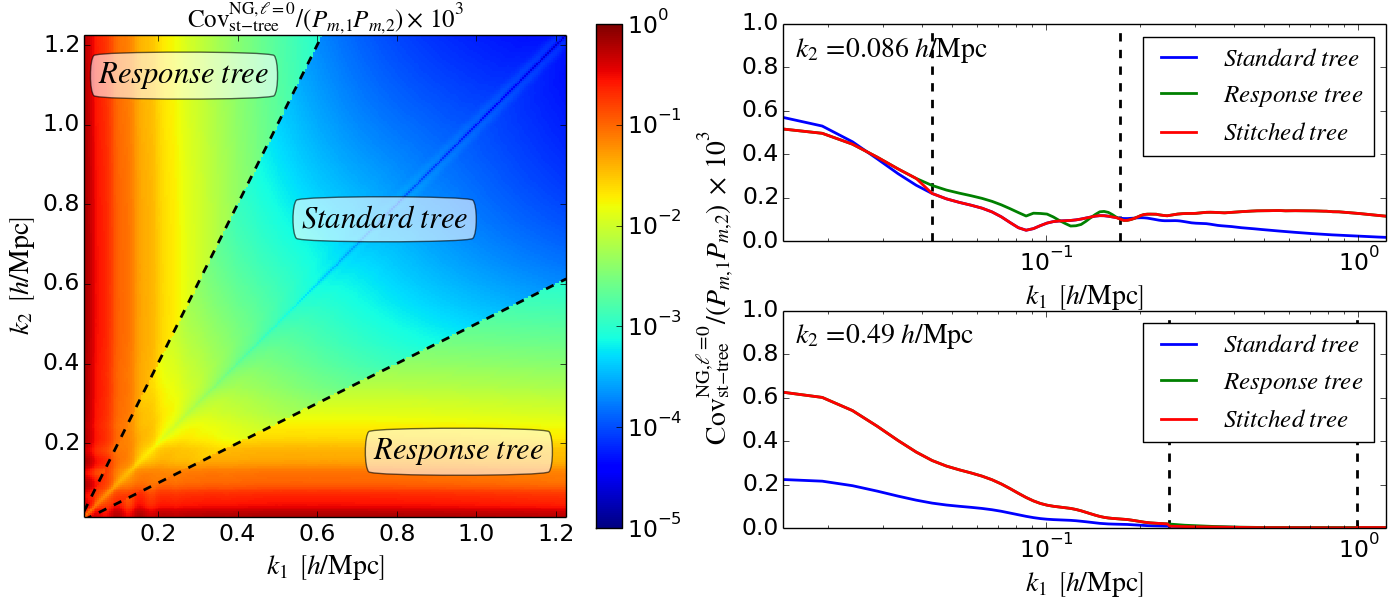}
	\caption{Non-Gaussian angle-averaged matter power spectrum covariance at tree level. The left panel shows the stitched tree-level covariance matrix of Eq.~(\ref{eq:stitched}) as a color plot. The dashed lines show $k_2 = f_\text{sq}k_1$, $k_2 = k_1/f_\text{sq}$, and draw the boundaries in $(k_1,k_2)$-space in which one either uses the standard tree-level or the response tree-level expressions, as labeled. The right panels show the stitched tree level covariance matrix (red) at two fixed $k_2$ values, as labeled. Also shown is the standard tree level result (blue) and the response tree-level expression (green). Results are shown for $V = 656.25\ h^{-3}{\rm Mpc}^3$. Further, $P_{m,i}\equiv P_m(k_i)$.}
\label{fig:tree}
\end{figure}

Our {\it stitched} tree-level result of Eq.~(\ref{eq:stitched}) is shown in Fig.~\ref{fig:tree}, for the angle-averaged case ($\ell = 0$ in Eq.~(\ref{eq:angleaver})). The matrix is shown as a color plot in the left panel. The two panels on the right show each a {\it slice} at fixed $k_2$ of the matrix on the left (red), as well as the standard (blue) and response (green) results. The upper right panel represents a slice at $k_2 < \knl$, while the lower right shows a slice at $k_2 > \knl$. The dashed black lines draw the boundary between the squeezed and non-squeezed limits. When both modes are smaller than $\knl \approx 0.3\kunit$ (at $z=0$), the color plot displays a smooth transition from the standard- to the response-based results, as it should be by definition. One does note, however,  that there are noticeable discontinuities at the junction between the two cases, when both $k$ values are above $\knl$. This is expected, as the standard tree-level result does not include any nonlinear corrections, while the response prediction does. Furthermore, even if, say, $k_2 \ll k_1$, but $k_2$ is of order $\knl$ or larger, then we do not expect the response tree-level result by itself to be a good description of the covariance. This is because loop contributions become non-negligible in that regime (cf.~$\mathcal{O}(k_\text{soft}^2/\knl^2)$ corrections in Eq.~(\ref{eq:resptreecov})). We will see below that, in this regime, our tree-level result is negligible compared to the contribution from 1-loop terms, and hence, the unphysical discontinuity in the stitched-tree-level contribution does not affect the much larger total covariance.

\section{Non-Gaussian covariance at the 1-loop level}\label{sec:NGcov1loop}

We now extend the calculation of the non-Gaussian part of the covariance by working at 1-loop level (cf.~Eq.~(\ref{eq:covngexp})). To do so, we introduce a new concept in large-scale structure perturbation theory which uses responses to describe the coupling between soft \emph{internal} loop momenta and hard external modes, thereby going beyond the so-far considered application of responses to describe the coupling of soft \emph{external} modes with hard external modes. 

The 1-loop covariance $\cov^{\rm NG}_{{\rm 1loop}}(k_1, k_2, \mu_{12})$ has contributions from nine types of diagrams (see e.g.~Fig.~4 of Ref.~\cite{2016PhRvD..93l3505B}). In the limit of soft loop momenta $p$, i.e. $p \ll k_1, k_2$, it can be shown that six of these diagrams are linear in $\Plin(p)$, i.e., they are of the form
\bq\label{eq:Pp}
\cov^{\rm NG}_{{\rm 1loop}, \Plin(p)}(k_1, k_2, \mu_{12}) \stackrel{p \ll k_1,k_2}{\propto} \int {\rm d}^3p\  \left[ \dots \right]\ \Plin(p)\Plin(k_i)\Plin(k_j)\Plin(k_k),
\eq
where $k_i,k_j,k_k$ are of the order of the external modes $k_1,k_2$.
Here and below, the dots in square brackets denote perturbation theory kernels $F_n$ that we do not write for brevity. 
On the other hand, the remaining three diagrams contain contributions that involve two powers of $\Plin(p)$, i.e.,
\bq\label{eq:Pp2}
\cov^{\rm NG}_{{\rm 1loop}, [\Plin(p)]^2}(k_1, k_2, \mu_{12}) \stackrel{p \ll k_1,k_2}{\propto} \int {\rm d}^3p\  \left[ \dots \right]\ [\Plin(p)]^2\Plin(k_i)\Plin(k_j)\,.
  \eq
These latter three diagrams can be represented exactly as a single diagram that involves two tree-level $\R_2$ vertices:
\ba\label{eq:R21loopdiag}
\raisebox{-0.0cm}{\includegraphicsbox[scale=0.8]{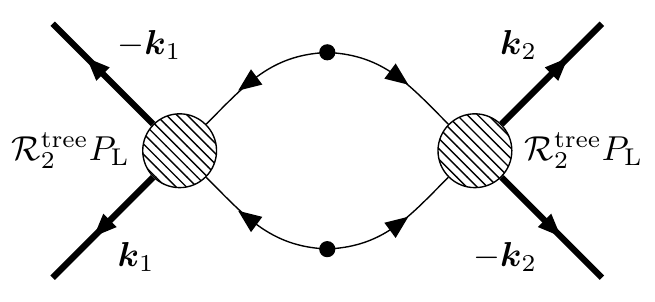}} &=& \left(\raisebox{-0.0cm}{\includegraphicsbox[scale=0.8]{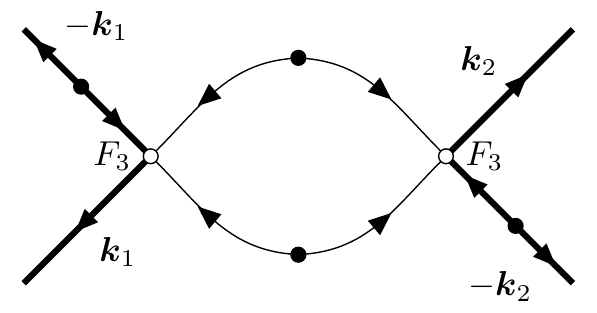}} + 3\ {\rm perms.}\right) \nonumber \\
&+& \left(\raisebox{-0.0cm}{\includegraphicsbox[scale=0.8]{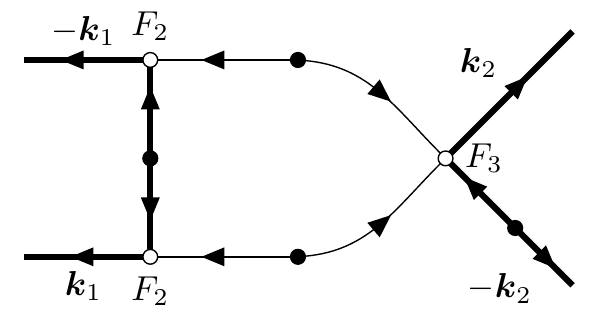}} +  3\ {\rm perms.}\right) \nonumber \\
&+& \left(\raisebox{-0.0cm}{\includegraphicsbox[scale=0.8]{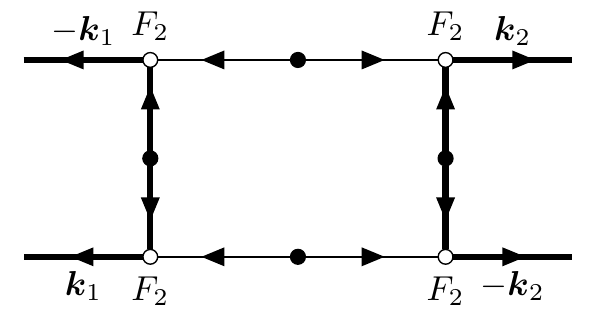}} +  1\ {\rm perm.}\right), \nonumber \\
\ea
where the role of the long-wavelength perturbations is played by internal loop momenta. The above equation is written in terms of perturbation theory kernels in Eq.~(\ref{eq:Cov1loop_tree}) in Appendix \ref{app:derivation}, where we demonstrate explicitly which standard 1-loop covariance terms are captured by the response approach. Based on the Feynman rules augmented with responses (cf.~Appendix \ref{app:feynman}), this represents a \emph{linking} of two diagrammatic representations of the tree-level response $\R_2^\text{tree}$ (cf.~Eq.~(\ref{eq:R2treediag})). As it was the case at tree level in the covariance, the generalization of the tree-level $\R_2$ to its simulation-calibrated expressions effectively extends the validity of the calculation of these 1-loop terms to nonlinear values of the hard modes $k_1, k_2$. A point that is worth emphasizing is that now both external modes can be of comparable size, i.e., this constitutes an application of the response formalism beyond the commonly used application to describe squeezed-limit correlation functions.

At this point, it may not be clear why capturing the terms in \refeq{Pp2} through responses yields a significant advantage. However, as we will comment on in the next subsection, the other 1-loop terms in \refeq{Pp} are suppressed relative to those in \refeq{Pp2} if $k_1,k_2$ are sufficiently large. Further, beyond the limit of soft loop momentum, the contributions from the opposite limit, i.e. $p \gg k_1,k_2$, are suppressed due to mass-momentum conservation. By expressing \refeq{Pp2} in terms of responses, one can therefore capture a substantial part of the covariance in $(k_1,k_2)$-space. Moreover, in the squeezed regime where $k_\text{soft} \ll k_\text{hard}$, additional response diagrams allow us to capture \emph{all} 1-loop terms that are leading order in $k_{\rm soft}/k_{\rm hard}$, although we do not calculate these in this paper. We return to this in \refsec{defnr}.

By the Feynman rules, the response diagram in \refeq{R21loopdiag} can then be written as
\ba\label{eq:Cov1loop}
&
\raisebox{-0.0cm}{\includegraphicsbox[scale=0.8]{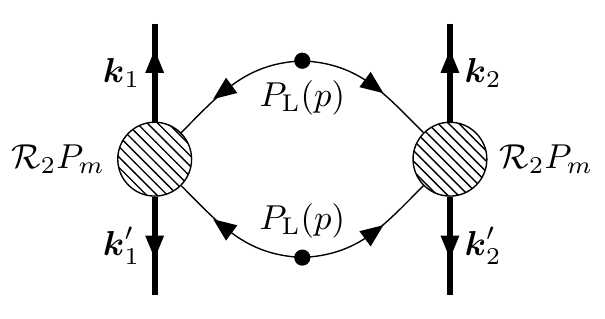}}
+(\text{perm.}) = \nonumber \\ 
& = 2 \int \frac{{\rm d}^3p}{(2\pi)^3}[\Plin(p)]^2 \R_2(k_1,\mu_1,-\mu_1,-1,1) P_m(k_1)\, \R_2(k_2,\mu_2,-\mu_2,-1,1) P_m(k_2) \nonumber \\ 
& = \frac{2P_m(k_1)P_m(k_2)}{(2\pi)^3}\Bigg[\int_0^{ p_\text{max}}p^2[\Plin(p)]^2{\rm d}p\Bigg] \nonumber \\
& \ \ \ \ \ \ \ \ \ \ \ \ \ \ \ \ \ \ \ \ \ \ \ \ \ \times \int_{-1}^1{\rm d}\mu_1\int_{0}^{2\pi}{\rm d}\varphi \R_2(k_1,\mu_1,-\mu_1,-1,1) \R_2(k_2,\mu_2,-\mu_2,-1,1),
\ea
where $\mu_1 = \vk_1\cdot\vp/(k_1 p)$, $\mu_2 = \vk_2\cdot\vp/(k_2 p)$ and the factor of 2 comes from the two possible ways of connecting the loops. In the above equation, we have implicitly fixed the direction of $\vk_1$, which means the polar integral of the loop momentum $\vp$ is done w.r.t.~$\vk_1$. 

We note for completeness that Eq.~(\ref{eq:Cov1loop}), as written, does not correctly describe cases where $\vk_1$ and $\vk_2$ are very nearly parallel such that $|\vp \pm \vk_1 \pm \vk_2| \approx p$. These terms can be straightforwardly included using responses as well (see Appendix \ref{app:derivation} for more details). However, after angle-averaging, their contribution to low-order multipoles is suppressed by $(p/k_i)^2$ and thus becomes negligible. We therefore do not include them in the main text. After performing the two angle integrals in Eq.~(\ref{eq:Cov1loop}) one arrives at
\bq\label{eq:cov1loop_res}
&& \cov^{{\rm NG}}_{\R\text{-1loop}} (k_1, k_2, \mu_{12}) = V^{-1}\frac{2P_m(k_1)P_m(k_2)}{(2\pi)^2}\Bigg[\int_0^{ p_\text{max}}p^2[\Plin(p)]^2{\rm d}p\Bigg]  \nonumber \\ 
&& \hspace*{2cm}\times \Bigg[2\A_1\A_2 + \frac25\B_1\B_2\P_2(\mu_{12}) + \frac25\big(\A_1\C_2+\A_2\C_1\big) + \frac{4}{35}\big(\B_1\C_2 + \B_2\C_1\big)\P_2(\mu_{12}) \nonumber \\ 
&&  \ \ \ \ \ \ \ \ \ \ \ \ \ \ \ \ \ \ \ \ \ \ \ \ \ \ \ \ \ \ \ \ \ \ \ \ \ \ \ \ \ \ \ \ \ \ \ \ \ \ \ \ \ \ \ \ \ \ \ \ \ \ \ \ + \frac{2}{35}\big(1+2\P_2(\mu_{12})^2\big)\C_1\C_2\Bigg],
\eq
where $\A_i \equiv \A(k_i)\ (i=1,2)$, and similarly for $\B_i$ and $\C_i$ (cf.~Eq.~(\ref{eq:PR2_anglecov}) for the definition of the $\A(k)$, $\B(k)$ and $\C(k)$ in terms of linear combinations of the response coefficients $R_O(k)$). 

A key issue to address before evaluating \refeq{cov1loop_res} concerns the value for the maximum loop momentum $p_\text{max}$. Our criterion to choose $p_\text{max}$ is based on the fact that Eq.~(\ref{eq:Cov1loop}) is only strictly valid if $p \ll k_1,k_2$, as well as $p < \knl$, otherwise, the blobs in \refeq{Cov1loop} would not correspond to response-type interactions. In this paper, we therefore choose the cutoff of the momentum integral to be
\be
p_\text{max} = \min \{ f_\text{sq} k_1,\, f_\text{sq} k_2,\, \knl \}\,.
\ee
Here, we employ the same fraction $f_\text{sq} = 0.5$ as used in our stitched tree-level result (cf.~Eq.~\ref{eq:stitched}), which assumes that the departure of the response prediction from the full 1-loop trispectrum away from the soft loop momenta limit scales similarly as in the tree-level case (cf.~Fig.~\ref{fig:fsq}). This is a reasonable assumption as the relevant interactions at both tree- and 1-loop levels are controlled by $\R_2$.

\subsection{Additional contributions to the 1-loop covariance}\label{sec:defnr}

The 1-loop contributions to the covariance that are not captured by \refeq{cov1loop_res} are of two types. One, in the limit of soft loop momentum, is the contribution from the six diagrams that are of the form of Eq.~(\ref{eq:Pp}). The other is the contribution from all 1-loop diagrams with loop momentum $p > p_\text{max}$. We comment on both contributions in turn below. We will conclude that  the non-response contributions are small compared to the response contributions everywhere except for $k_1, k_2 \sim 0.1-0.3\iMpch$. These missing terms can nonetheless be included with a stitching procedure analogous to that performed for the tree-level covariance in \refsec{NGcovtree}.

Let us first consider \refeq{Pp}. The relative size of these non-response terms  compared to the response-type ones (cf.~Eq.~(\ref{eq:Pp2})) can be roughly estimated by
\bq\label{eq:ratioest}
\frac{\Plin(k_\text{soft})\int_0^{p_\text{max}}p^2\Plin(p){\rm d}p}{\int_0^{p_\text{max}}p^2[\Plin(p)]^2{\rm d}p},
\eq
where $k_\text{soft} \gtrsim 0.1\kunit$. In the numerator, it makes sense to use the power spectrum evaluated at $k_\text{soft}$ because $\Plin(k_\text{soft}) > \Plin(k_\text{hard})$ in the regime of interest, so that \refeq{ratioest} captures the most relevant terms (we are setting the perturbation theory kernels to unity for this estimate). For our choice of $p_\text{max}$, we have for the above ratio
\be
\mbox{\refeq{ratioest}} \approx \{ 0.27,\  0.11, \  0.02\}
\quad\mbox{for}\quad 
k_\text{soft} = \{ 0.1,\   0.3,\   1\} \kunit\,,
\ee
respectively. This indicates that, at the transition from the linear to the nonlinear regime in soft external momenta, $k_{\rm soft} \sim 0.1 - 0.3\kunit$, there are 1-loop terms that are sizeable, but that are not of the type of Eq.~(\ref{eq:Cov1loop}). When both $k_1, k_2 \sim 0.1 - 0.3\kunit$ these missing terms can be calculated with standard perturbation theory; if, on the other hand, $k_{\rm soft} \sim 0.1 - 0.3\kunit$ but $k_{\rm hard} \gg k_{\rm soft}$, then the missing terms can be evaluated by combining standard perturbation theory and response vertices in the same diagram (see e.g.~Eq.~(2.12) of Ref.~\cite{paper1}).  Note that for values of $k_\text{soft} \lesssim 0.1\kunit$, the contribution from the 1-loop term is small, and as a result, it is numerically irrelevant whether the 1-loop contribution is accurate.

We now turn to the second missing 1-loop part, namely the contribution from loop momenta with $p > p_\text{max}$.  Consider a loop momentum $p \gg k_1, k_2$. This  corresponds to mode-coupling interactions in which hard ingoing momenta combine to form outgoing soft momenta. These types of couplings are suppressed by momentum and mass conservation (see Appendix~B of Ref.~\cite{abolhasani/mirbabayi/pajer:2016} for a more detailed discussion). Specifically, the perturbation theory kernels in this limit scale as $(k_i/p)^2$ ($i=1,2$), and as a result, the loop integrals in this regime contribute negligibly to the total covariance. This, combined with the fact that the response-type terms dominate for $k_\text{soft} \gtrsim \knl$, restricts the contributions from loop momenta $p > p_\text{max}$ to the regime of $k_1, k_2 \sim 0.1 - 0.3\kunit$, as well.

We shall return to the importance of these missing contributions below, as we analyze the results of our calculation. We stress that the inadequacy of the response-based approach to correctly describe the 1-loop covariance for $k_1, k_2 \sim 0.1 - 0.3\kunit$ can be circumvented by a {\it stitching} to the standard 1-loop calculation (see Ref.~\cite{2016PhRvD..93l3505B} for the complete expressions), similar to that implemented in the last section for the tree-level covariance. We leave such a {\it stitching} at the 1-loop level (as well as the inclusion of other terms important for $k_{\rm soft} \sim 0.1 - 0.3\kunit$ but $k_{\rm hard} \gg k_{\rm soft}$)  for future work.

\subsection{Estimate of higher-loop contributions}\label{sec:defhl}

We have argued above that, for sufficiently high $k$, the 1-loop contribution to the covariance is dominated by response-type terms. This does not address, however, the issue of the relevance of higher loops on these scales, which we consider now. 

Let us consider the 2-loop contribution to the covariance in the response approach. This corresponds to a single diagram that is the $n=3$ generalization of Eq.~(\ref{eq:Cov1loop}): 
\ba
&
\raisebox{-0.0cm}{\includegraphicsbox[scale=0.8]{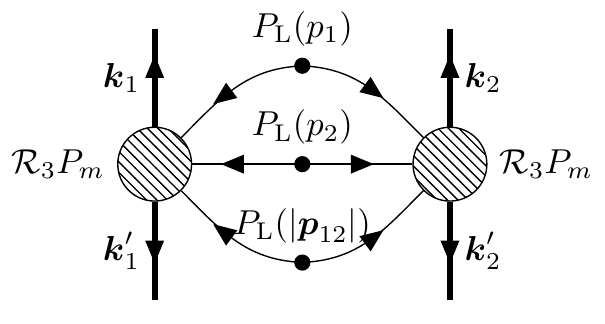}}
+ (\text{perm.}) = \vs
&\hspace*{0cm}  = 6 P_m(k_1)P_m(k_2) \int_{\vp_1} \int_{\vp_2} \Plin(p_1) \Plin(p_2) \Plin(|\vp_{12}|) \R_3(k_1, \cdots) \R_3(k_2, \cdots) \nonumber \\
& \ \ \ \ \ \ \ \ \ \ \ \ \ \ \ \ \ \ \ \ \ \ \ \ \ \ \ \ \ \ \ \ \ \ \ \ \ \ \ \ \times (2\pi)^3 \d_D(\vk_1+\vk_1'+\vk_2+\vk_2'),
\label{eq:Cov2loop}
\ea
where the dots in the arguments of $\R_3$ represent all the angles involved (omitted for brevity) and the factor of 6 accounts for the permutations of the internal loop momenta. Further, $\int_{\vp} \equiv \int {\rm d}^3p/(2\pi)^3$. An order-of-magnitude estimate of the relative size of this 2-loop contribution to that of Eq.~(\ref{eq:Cov1loop}) can be written as
\be
\frac{[\text{Response 2-loop}]}{[\text{Response 1-loop}]} \sim
\frac{6}{2} \left(\frac{\<\R_3\>_{\hat{\vp}_1,\hat{\vp}_2}}{\<\R_2\>_{\hat{\vp}}}\right)^2 \s_{p_\text{max}}^2\,,
\ee
where
\be
\s_{p_\text{max}}^2 \equiv \frac1{2\pi^2}\int_0^{p_\text{max}}{\rm d}p\ p^2  \Plin(p)
\ee
is the variance of the density field up to the cutoff $p_\text{max}$ employed in the loop integrals, and $\< \>_{\hat{\vp}_i}$ denotes the angle-average over the responses in Eqs.~(\ref{eq:Cov1loop}) and (\ref{eq:Cov2loop}). As a very rough estimate, we now assume that only the isotropic response coefficients $R_n/n!$ remain after these angle averages. Note, for instance, that comparing Eqs.~(\ref{eq:Cov1loop}) and (\ref{eq:cov1loop_res}) shows that for $\R_2$ this is not really correct. Keeping this caveat in mind, we obtain
\be
\frac{[\text{Response 2-loop}]}{[\text{Response 1-loop}]} \sim
\frac{1}{3} \left(\frac{R_3}{R_2}\right)^2 \s_{p_\text{max}}^2\,.
\ee
By continuing this reasoning to higher loops one obtains
\bq
\frac{[\text{Response $n$-loop}]}{[\text{Response 1-loop}]} \sim 2 [(n+1)!]^{-1} \left(\frac{R_{n+1}}{R_2}\right)^2 \s_{p_\text{max}}^{2(n-1)}.
\eq
This order-of-magnitude estimate leaves open the possibility that higher-loop terms in the response approach contribute non-negligibly to the total covariance if $\s_{p_\text{max}} \gtrsim 1$, which corresponds to $p_\text{max} \gtrsim \knl$. The importance of higher-loop response terms and how their contribution scales with $n$ is dependent also on the details of the shape of the $\R_n$, or more accurately, on the specific angle-averages that characterize the corresponding diagrams. These higher-order response functions have however never been fully derived, which prevents us from drawing decisive conclusions here. Interestingly, Ref.~\cite{response} found that the \emph{Eulerian} isotropic response coefficients $R_n^E(k)$, which measure the response of the power spectrum to evolved isotropic modes, are rapidly suppressed numerically at higher orders $n \geq 2$. 

We will return to the potential importance of higher-loop terms as we analyze the results of our covariance calculations below. We stress that, for given $k_1, k_2$, the importance of higher-loop terms should progressively decrease in order to render the response approach to the covariance well-defined and predictive. This highly relevant open issue is left for future investigation.

\section{Comparison with simulations: angle-averaged case}\label{sec:monosims}

We now assess the performance of the matter covariance expressions developed in the previous sections by comparing them to estimates from N-body simulations. In particular, we use the results of Ref.~\cite{blot2015}, who estimated the covariance matrix of the matter power spectrum by cross-correlating the angle-averaged power spectra from more than 12000 simulation boxes with volume $V = [656.25\ h^{-1} {\rm Mpc}]^3$. In this section, we therefore consider only the monopole (angle-averaged) part of our covariance expressions ($\ell=0$ in Eq.~(\ref{eq:angleaver})). In Ref.~\cite{blot2015}, the authors presented results from two sets of simulations: one called Set A, which consists of $12288$ realizations with $N_p = 256^3$ matter tracer particles; and one called Set B, which is made up of a lower number of realizations, $96$, but at higher resolution $N_p = 1024^3$. Apart from providing an independent estimate of the covariance, the diagonal components of the covariance estimated in Set B are used to derive a correction for the power spectrum as well as the covariance measured from Set A for mass resolution effects (see Ref.~\cite{blot2015} for details). In this paper, we show the covariance matrices of Ref.~\cite{blot2015} estimated from the spectra of Set B and the spectra of Set A after this correction is applied. Note that the diagonal elements of the covariance of both sets thus agree by definition, and that the correction applied cancels out when considering the correlation coefficient Eq.~(\ref{eq:corrcoeff}).

The cosmological parameters of the simulations of Ref.~\cite{blot2015} (cf.~end of Sec.~\ref{sec:intro}) are almost the same as those used in Ref.~\cite{response} to measure the isotropic response coefficients $R_1(k)$ and $R_2(k)$, which is why we choose these data to compare our results with. We note that our formalism holds generically for any quintessence-type cosmology, provided the corresponding power spectrum responses are known. Other recent estimates of the covariance matrix using simulations include those of Ref.~\cite{li/hu/takada}, who account for the super-sample covariance term, as well as those in Ref.~\cite{2017arXiv170105690K}, which were obtained using over 15000 simulations (see also Refs.~\cite{2006MNRAS.371.1188H, 2009ApJ...700..479T, joachimpen2011, 2011ApJ...734...76S}).

Before discussing the detailed comparison, we make some cautionary remarks regarding simulation measurements of the power spectrum covariance. As any measurement, they in general have statistical and systematic errors. The statistical errors are due to the finite number of realizations, or total volume, of the simulations. On large scales (small wavenumber) the statistical error on the simulation measurements is dominated by the limited number of modes sampled. For a total simulated volume $V_t$, this number is given by $k^3 V_t/(2\pi)^3$, and hence it is smallest (largest statistical error) for modes close to the fundamental mode of the individual boxes $k_\text{fund} = 2\pi/L_\text{box}$ (where $L_\text{box}$ is the box size). On nonlinear scales and thus higher wavenumber, these \emph{sample variance} effects become smaller, but the precise error becomes harder to quantify because of mode coupling that effectively correlates the statistical error of the covariance across different wavenumbers. The systematic errors of N-body simulations include the finite resolution due to the number of particles, the subtraction of particle shot noise, and transients from the initial conditions. The first two contributions are expected to be most significant on the smallest scales. Quantifying the systematic error on the estimated power spectrum covariance without numerical convergence tests is very difficult. Further, by definition, simulation-based estimates of the covariance matrix lack the contribution from modes with $k < k_\text{fund} = 2\pi/L_\text{box}$, which can also be seen as systematic error. Note however that, as shown in Ref.~\cite{li/hu/takada}, these can be included at leading order through the super-sample-variance contribution.

Strictly, for comparison with these simulations, one should also include a minimum value $p_\text{min} = k_\text{fund}$ in the loop integrals of our calculation. However, for $L_\text{box} \sim 650\Mpch$, we have found that this makes an entirely negligible numerical difference.

Due to the difficulty of obtaining reliable error estimates on the simulation-based covariance, we will mainly discuss the comparison of our covariance prediction to simulation results in the context of the known deficiencies of the former. As discussed in Secs.~\ref{sec:defnr} and \ref{sec:defhl}, these are non-response-type terms on quasi-linear scales, and higher-loop response terms on fully nonlinear scales. We deliberately avoid quantifying the exact level of agreement between theory and simulations, since it could be misleading given the above mentioned difficulties in estimating the error on the latter.

\subsection{Comparison at $z=0$}\label{sec:z0}

\begin{figure}[t]
	\centering
	\includegraphics[width=0.9\textwidth]{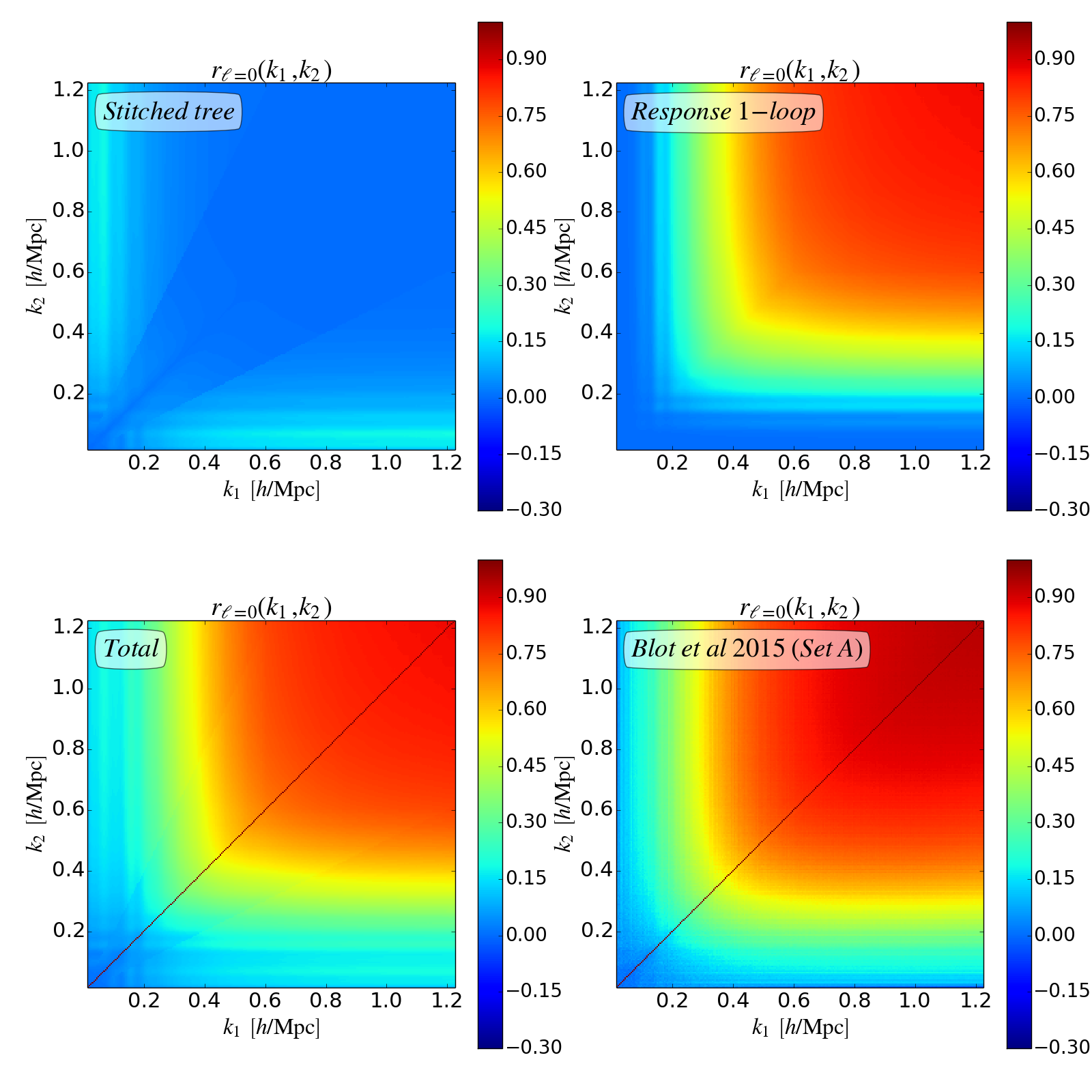}
	\caption{Correlation coefficient of the angle-averaged matter covariance, $r_{\ell = 0}(k_1, k_2)$ at $z=0$ (cf.~Eq.~(\ref{eq:corrcoeff})). The four panels display the contribution from the stitched tree level and response 1-loop parts (upper panels), as well as their summed result (together with the Gaussian contribution) and the estimates from the simulations of Set A of Ref.~\cite{blot2015} (lower panels), as labeled. In the upper panels, the matrix used in the denominator of Eq.~(\ref{eq:corrcoeff}) is the total covariance matrix, such that the lower left panel is obtained by summing the two upper ones (in addition to the Gaussian contribution).}
\label{fig:map}
\end{figure}

\begin{figure}[t]
  \vspace{1cm}
	\centering
	\includegraphics[width=\textwidth]{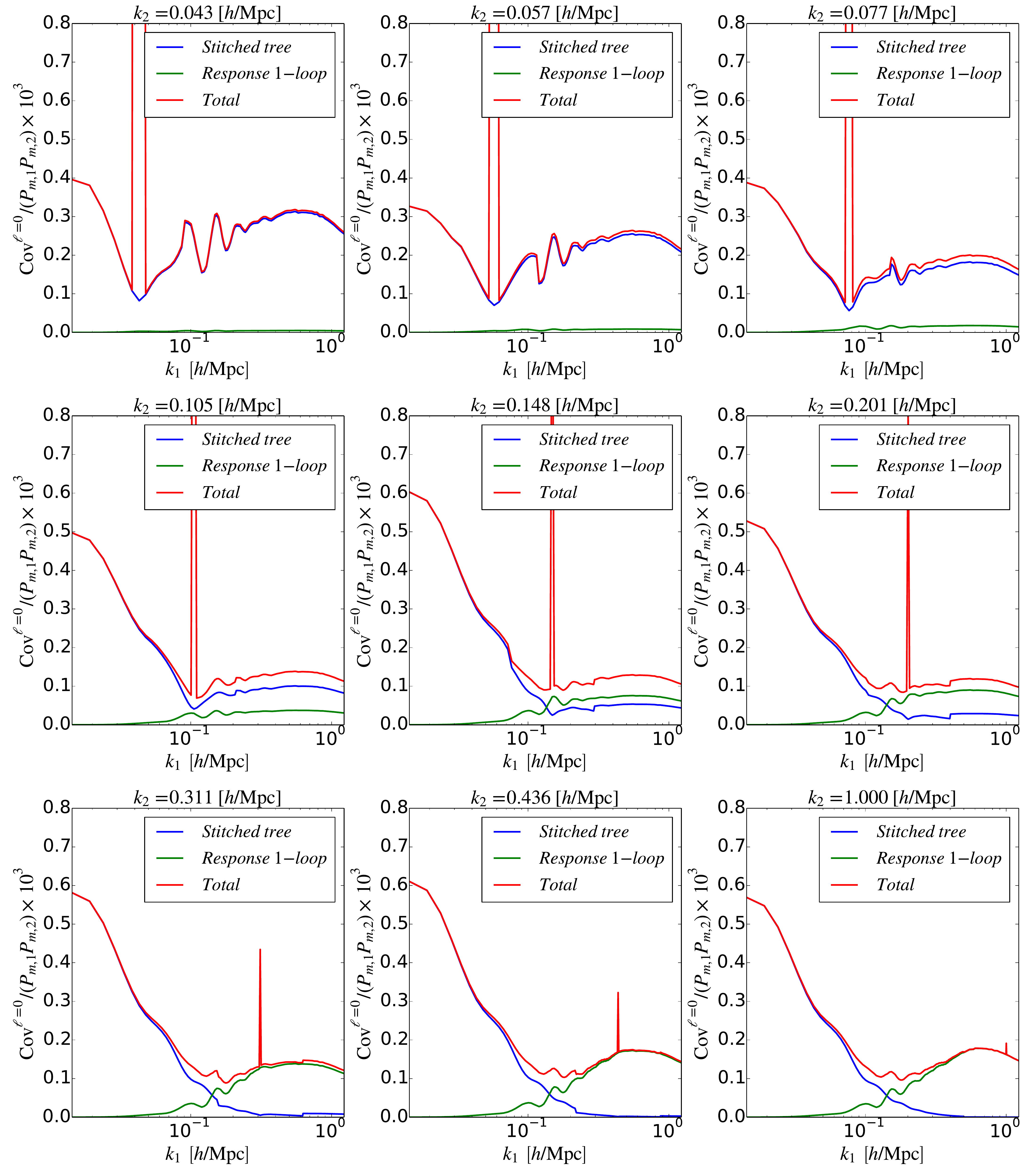}
  \vspace{1cm}
	\caption{Covariance matrix as a function of $k_1$, for fixed values of $k_2$ (as indicated in the title of each panel) at $z=0$. Each panel shows our stitched tree-level and response 1-loop results, as well as their sum (including also the Gaussian term, visible as the sharp spikes at $k_1=k_2$), as labeled.}
\label{fig:slices}
\end{figure}

The color plots in Figure \ref{fig:map} show the correlation coefficient of the angle-averaged matter power spectrum evaluated at two different wavenumbers $k_1, k_2$, which is defined as
\bq\label{eq:corrcoeff}
r_{\ell = 0}(k_1, k_2) = \frac{\cov^{\ell = 0}(k_1, k_2)}{\sqrt{\cov^{\ell = 0}(k_1, k_1)\cov^{\ell = 0}(k_2, k_2)}}.
\eq
$r_\ell(k_1,k_2)$ can take on values between $-1$ and $1$. The upper panels show the contribution from the stitched tree-level non-Gaussian term (cf.~Eq.~(\ref{eq:stitched}); upper left) and response 1-loop result (cf.~Eq.~(\ref{eq:cov1loop_res}); upper right). The lower left panel shows the total prediction for the angle-averaged covariance, which includes the Gaussian diagonal contribution as well. In the upper panels and in the lower left panel, we use the total covariance in the denominator of $r_{\ell = 0}(k_1, k_2)$, i.e., the lower left panel is obtained by summing the two upper panels and adding the Gaussian contribution. Figure \ref{fig:slices} shows instead a few representative \emph{slices} at constant $k_2$ of the covariance matrix $\cov^{\ell = 0}(k_1, k_2)$ (not the correlation coefficient).

The upper panels of Fig.~\ref{fig:map} and the panels in Fig.~\ref{fig:slices} are pedagogical in that they illustrate the kinematical regimes in which the tree level and the 1-loop terms contribute most. In particular, the tree-level result dominates when at least one of the modes is $\lesssim 0.1\kunit$. On the other hand, when both modes are $\gtrsim 0.1\kunit$, then most of the contribution comes from the 1-loop term (recall that we do not include the non-response-type loop contribution). The various panels of Fig.~\ref{fig:slices} help to visualize the gradual increase in importance of the 1-loop term as $k_2$ becomes larger. For instance, in the upper left panel for $k_2 = 0.043\kunit$, the 1-loop contribution is fairly small and almost all of the non-diagonal covariance is captured at tree level (blue line). As $k_2$ increases however (left to right, top to bottom), the tree-level result becomes progressively smaller at high $k_1$, and is complemented by the growing contribution of the 1-loop term (green line). In light of the relative importance of tree-level and 1-loop contributions, the sharp discontinuities at high $k$ between the two branches of the stitched tree-level result of Eq.~(\ref{eq:stitched}), as well as the extrapolation of the tree-level response to the case of $k_\text{soft} \gtrsim \knl$ are not affecting the total covariance in this regime because the entire stitched tree-level contribution is a small part of the total result. 

\begin{figure}[t]
	\centering
	\includegraphics[width=\textwidth]{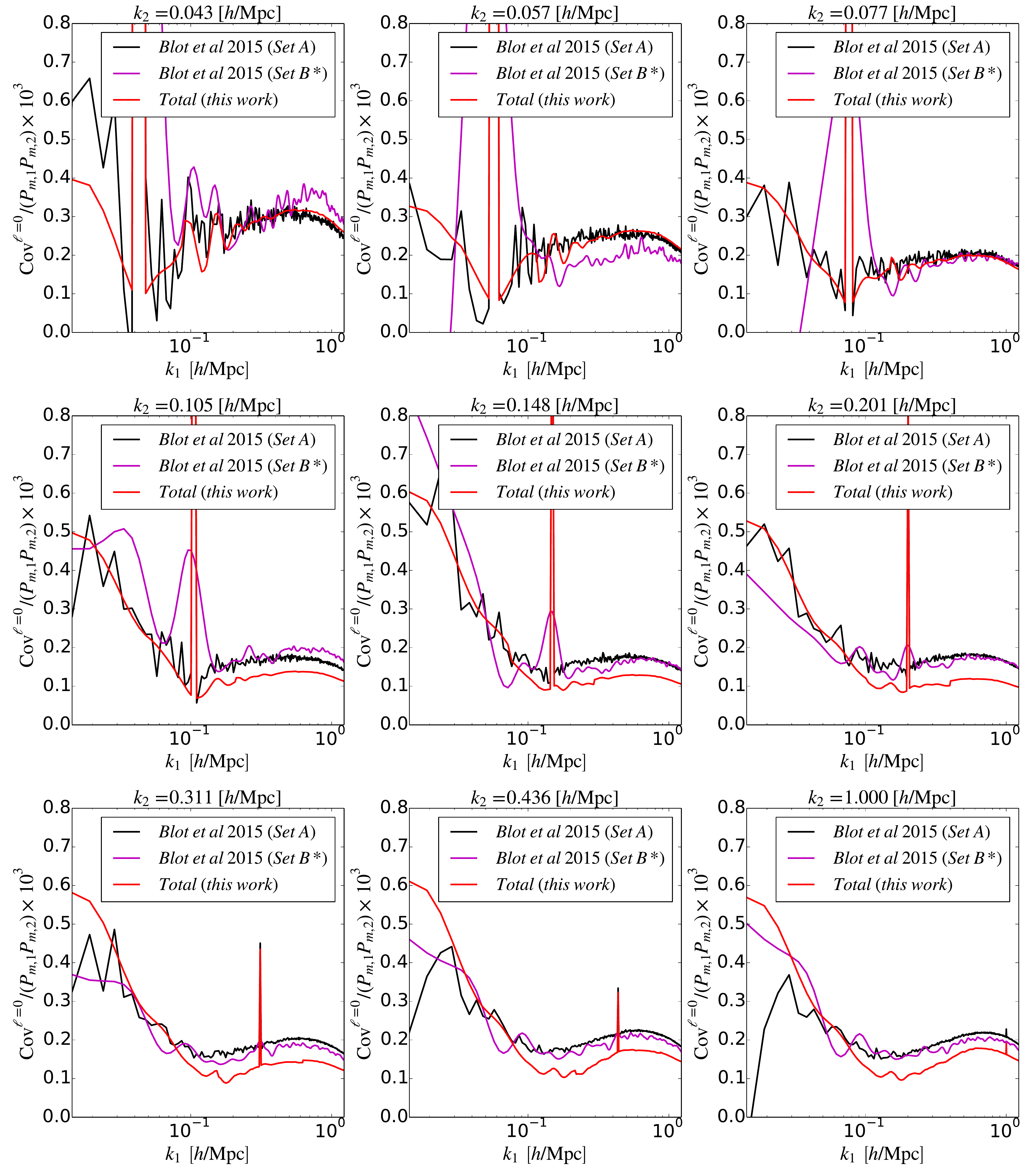}
	\caption{Covariance matrix as a function of $k_1$, for fixed values of $k_2$ (as indicated in the title of each panel) at $z=0$. Each panel shows the simulation results of Ref.~\cite{blot2015}, as well as the result from our calculation, as labeled. The $k_2$ values are the same as in Fig.~\ref{fig:slices}. The discrepancy between theory and simulations for $k_1 \lesssim 0.03\kunit$ in the lower three panels can be attributed to insufficient volume of the simulations to sample these large modes. In the labels of the y-axis, $P_{m,i}\equiv P_m(k_i)$, which we evaluate using the {\sc Coyote} emulator. The $^*$ in the label of the simulation Set B indicates that the covariance matrix was smoothed with a Gaussian kernel to reduce the noise and facilitate visualization of the \emph{trends} in the measurements.}
\label{fig:slicessims}
\end{figure}

The lower right panel of Fig.~\ref{fig:map} shows $r_{\ell=0}(k_1, k_2)$ from the simulation Set A of Ref.~\cite{blot2015}. The visual comparison to our prediction does not reveal strong differences in either shape or overall amplitude. A more detailed comparison with simulations is shown in Fig.~\ref{fig:slicessims}, where we show our total prediction along with the simulation Set A and Set B results of Ref.~\cite{blot2015}. Up to the approximation employed in the extrapolation of the anisotropic response coefficients $R_O(k)$ (cf.~Appendix \ref{app:ro}), our calculation is guaranteed to capture the total covariance if the soft mode is sufficiently linear, $k_\text{soft} \ll \knl \approx 0.3\kunit$. An interesting application of our calculation in this regime is therefore to test simulation-based estimates of the covariance matrix for systematic errors. Indeed, both simulation sets are in relatively good agreement with our calculation whenever $k_\text{soft} \ll \knl$; this includes roughly the whole $k_1$ range in the upper three panels, as well as the low-$k_1$ parts of the lower six panels in Fig.~\ref{fig:slicessims}. The differences between Set A and Set B are likely to be mostly caused by the larger statistical uncertainties in Set B due to the smaller volume covered. Further, the departures seen in the simulations for large-scale modes, $k_1 \lesssim 0.03\kunit$ (noticeable in the lower three panels of Fig.~\ref{fig:slicessims}), are likely to be due to insufficient sampling of these modes by both sets of simulations, as we noted already in the beginning of this section.

The lower six panels of Fig.~\ref{fig:slicessims} correspond to slices with $k_2 > 0.1\kunit$, in which one notes that our calculation falls short of describing completely the simulation measurements for $k_1 \gtrsim 0.1\kunit$. As we have discussed in Secs.~\ref{sec:defnr} and \ref{sec:defhl}, there are two types of terms that are expected to contribute non-negligibly in this regime, presumably accounting for a large fraction of the observed difference between theory and simulations. One corresponds to 1-loop diagrams that cannot be brought into the form of Eq.~(\ref{eq:Cov1loop}) and that can contribute sizeably when $k_1, k_2 \sim 0.1-0.3\kunit$. The inclusion of these terms via a stitching of standard- and response-based expressions should render the whole calculation fully predictive in this regime. In regimes in which $k_{\rm soft}\sim 0.1-0.3\kunit$ and $k_{\rm hard} > \knl$ (e.g.~$k_1 = 0.2\kunit$ in the lower right panel of Fig.~\ref{fig:slicessims}), the missing terms can also be added, and in fact, using responses to describe the interactions that involve $k_{\rm hard}$. These terms, however, are not expected to play a major role in cases when $k_1, k_2 > 0.3\kunit$. In this kinematic regime on the other hand, one expects that 2- and higher-loop terms, which themselves are dominated by response-type terms in this regime (cf.~Eq.~(\ref{eq:Cov2loop})), can account for the missing contribution. An important requirement for the response approach to remain predictive when $k_1, k_2 > 0.3\kunit$ is, therefore, that higher-loop response-type contributions become progressively smaller. For the time being, we cannot provide a conclusive answer on the exact relative size of higher-loop response terms, and defer that to future work.

\begin{figure}[t]
	\centering
	\includegraphics[scale=0.55]{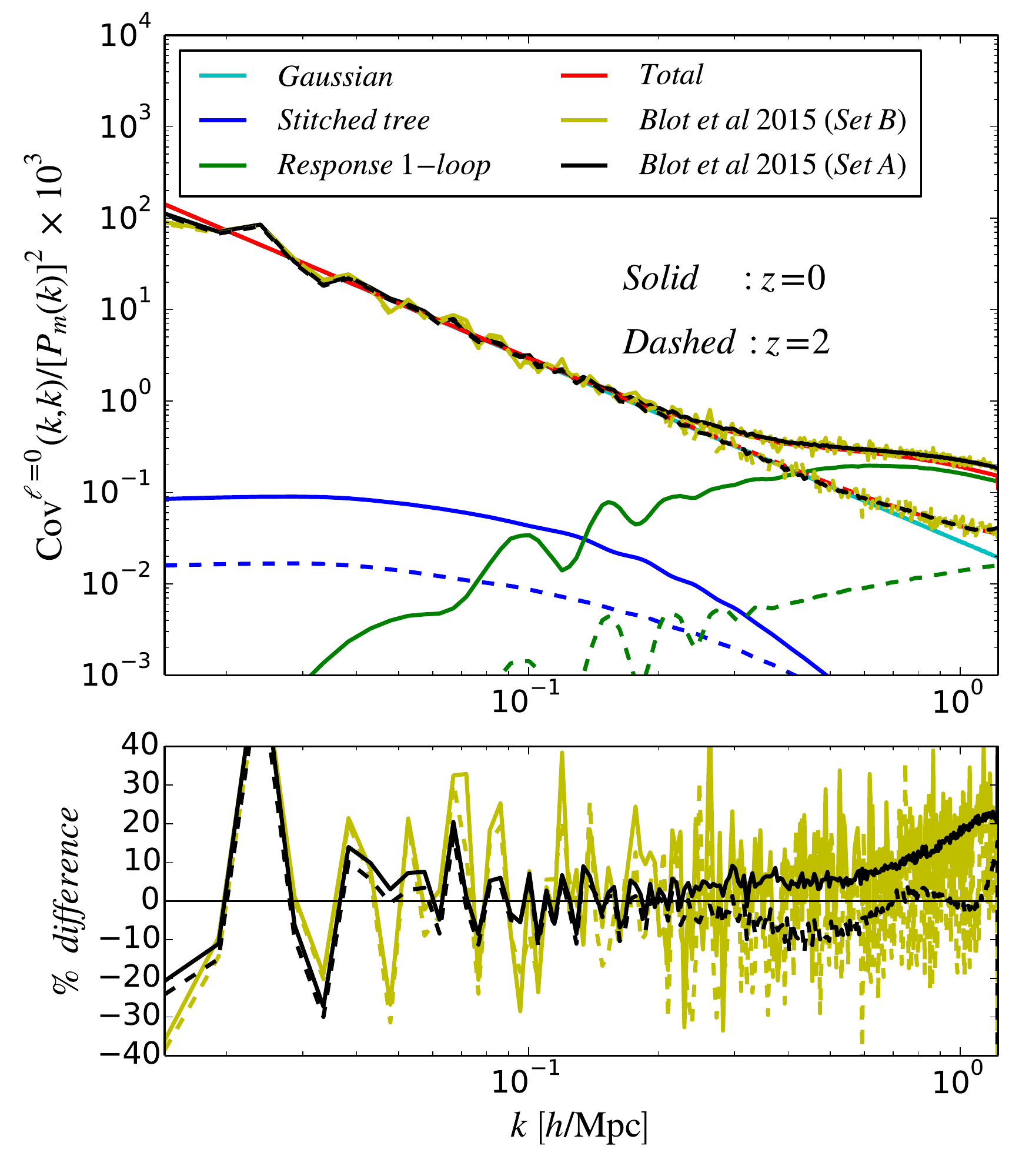}
	\caption{Diagonal of the covariance matrix at $z=0$ (solid) and $z=2$ (dashed). The upper panel shows the simulation results of Ref.~\cite{blot2015} and the result of our calculation, together with its Gaussian, stitched tree-level and response 1-loop parts, as labeled. The lower panel shows the fractional deviation of the simulation results to our calculation. With the normalization adopted for the $y$-axis, the Gaussian line is independent of redshift. The spectra in the denominator is the same for all curves and is evaluated using the {\sc Coyote} emulator.}
\label{fig:diagonal}
\end{figure}

It is also instructive to compare predictions for the diagonal of the covariance matrix ($k_1=k_2$), which is shown in Fig.~\ref{fig:diagonal} (solid lines for $z=0$). On large scales, this is dominated by the Gaussian contribution, and the tree-level non-Gaussian contribution is subdominant at all $k$ values. The 1-loop contribution only starts to become important for $k \gtrsim 0.3\kunit$. As a result, for $k \lesssim 0.1\kunit$, there is good agreement between our calculation and the simulations (up to noise), but this is unsurprising because here the result is set by the trivial and well understood Gaussian contribution. For $k \gtrsim 0.3\kunit$, the simulation results at $z=0$ (solid lines) have a higher amplitude than our calculation (better discernible in the less noisy Set A), but, as already mentioned above, this is a regime in which higher-loop terms are expected to contribute non-negligibly, and hence potentially reduce the gap between theory and simulations.

\subsection{Comparison at $z=2$}\label{sec:z2}

\begin{figure}
	\centering
	\includegraphics[width=\textwidth]{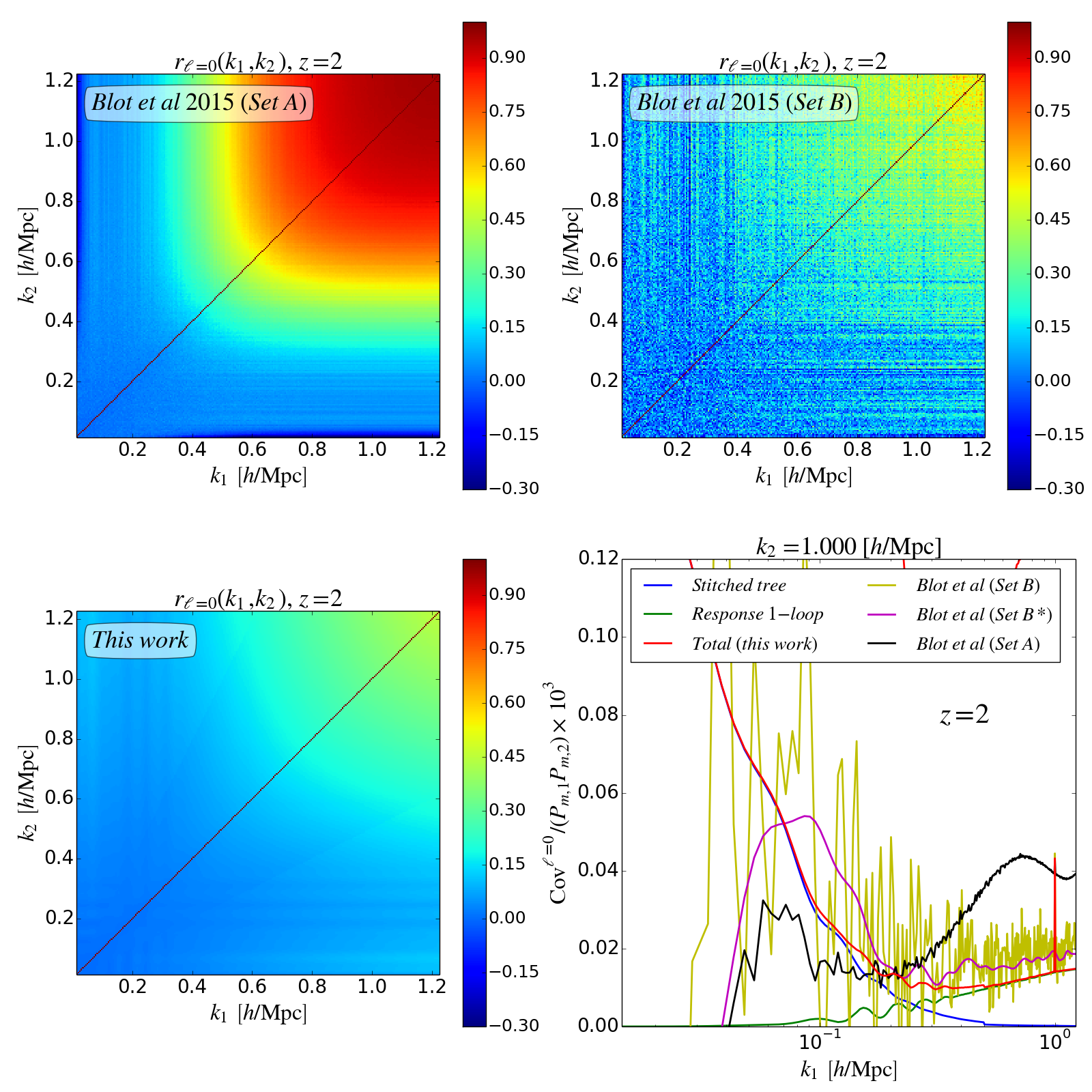}
	\caption{The upper panels show the $z=2$ correlation coefficient estimates from Ref.~\cite{blot2015} using their two sets of simulations: Set A, which comprises 12288 lower resolution $N_p = 256^3$ simulations; and Set B, which is made of $96$ higher resolution $N_p = 1024^3$ simulations, where $N_p$ is the N-body tracer particle number. The lower left panel shows the corresponding result from our model. The lower right panel shows the $z=2$ covariance matrix as a function of $k_1$, for fixed $k_2 = 1\kunit$ ($P_{m,i}\equiv P_m(k_i)$ in the label of the y-axis is evaluated using the {\sc Coyote} emulator). The $^*$ in the label of the simulation Set B indicates that the covariance matrix was smoothed with a Gaussian kernel to reduce the noise and facilitate visualization of the \emph{trends} in the measurements.}
\label{fig:issuesz2}
\end{figure}

Our prediction for the covariance can be straightforwardly applied to other redshifts as well. To do so, one should use the response coefficients $R_O(k)$ at the desired redshift.  This includes 1) using the measured isotropic responses from simulations and adjusting appropriately the nonlinear extrapolation of the anisotropic ones (cf.~Appendix \ref{app:ro}); 2) evaluate all spectra at the desired redshift; 3) and using the corresponding value of the nonlinear scale $\knl$ in setting $p_\text{max}$, which increases with redshift.

The color plots in Fig.~\ref{fig:issuesz2} show the correlation coefficient measured from both simulation sets of Ref.~\cite{blot2015} and that of our calculation at $z=2$, as labeled. The correlation coefficients measured from the two simulation sets are noticeably different at this redshift. In Ref.~\cite{blot2015}, this is attributed to the fact that the total volume of simulations in Set B (96 realizations) is not sufficient to appropriately sample the covariance. Note however that  the results from the two simulation sets are in much better agreement at $z=0$, as we have seen above. Interestingly, our prediction agrees markedly better with the result from Set B. This becomes clearer from the lower right panel of Fig.~\ref{fig:issuesz2}, which shows the slice of the covariance matrix at constant $k_2 = 1\kunit$ (the same as the lower right panel of Fig.~\ref{fig:slices}, but for $z=2$). Our calculation underpredicts both simulation set results, but the level of disagreement between our model and Set A is significantly larger than that with Set B. In fact, at $z=2$ the performance of our calculation in reproducing the results from Set B is comparable to the performance of the same in reproducing the results from both Set A and Set B at $z=0$. The lower volume of the higher-resolution simulation Set B unfortunately prevents us from drawing robust conclusions on the significance of its better agreement (compared to Set A) with our theoretical prediction.

This picture becomes different if one focuses only on the diagonal of the covariance matrix at $z=2$, which is shown by the dashed lines in Fig.~\ref{fig:diagonal}. As mentioned at the beginning of this section, the correction of Set A obtained by matching the diagonal elements to Set B ensures that the two agree on the diagonal, within the noise of the smaller Set B. They both agree well with our prediction. Note that this level of agreement for $k \gtrsim 0.6 \kunit$ depends quite crucially on the contribution from the 1-loop term.

\section{Angular dependence of the matter power spectrum covariance}\label{sec:angles}

\begin{figure}
	\centering
	\includegraphics[width=\textwidth]{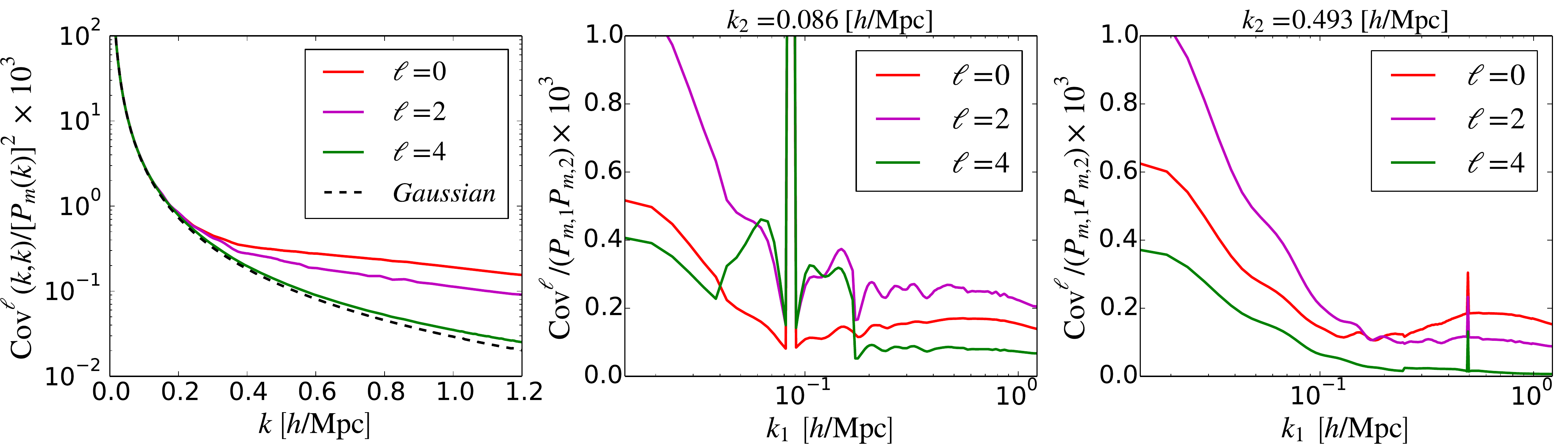}
	\caption{The left panel shows the diagonal of the multipoles $\ell=0, 2, 4$ of the total covariance matrix given by our prediction, as labeled. The dashed line indicates the Gaussian contribution to the covariance. The middle and right panels show the same three multipoles as a function of $k_1$, for two fixed values of $k_2$, as labeled. All results shown correspond to $z=0$, $V = 656.25\ h^{-3} {\rm Mpc}^3$ and $P_{m,i}\equiv P_m(k_i)$.}
\label{fig:angle}
\end{figure}

We now go beyond the case of the angle-averaged non-Gaussian covariance and analyze its angular dependence, which can be organized into Legendre multipoles according to Eq.~(\ref{eq:angleaver}). For the case of the response-based contributions, the well defined analytical dependence on $\mu_{12}$ allows for all multipoles to be evaluated analytically (cf.~Appendix \ref{app:analy}). The angular dependence of the standard perturbation theory contributions (the tree-level one in our case) is more cumbersome and we perform the angle-averages numerically.

Figure \ref{fig:angle} displays a few predictions at $z=0$ for the quadrupole ($\ell=2$) and hexadecupole ($\ell=4$), as well as the monopole ($\ell=0$) case studied more extensively in the previous section. The left panel shows the diagonal of these three multipoles. One notes that there is a hierarchy between the diagonal of these terms, with the monopole being the largest and the hexadecupole the smallest. The case depicted by the dashed black line corresponds to the Gaussian result of Eq.~(\ref{eq:gauss_mono}). All multipoles of the Gaussian contribution have this form. On scales $k\lesssim 0.3\kunit$, all multipoles match the Gaussian result because on these scales the fractional size of the non-Gaussian contribution to the diagonal is negligible.

The left two panels of Fig.~\ref{fig:angle} display {\it slices} at constant $k_2$ of the multipoles, which show that the hierarchy displayed along the diagonal does not necessarily hold for off-diagonal terms of the covariance. We note also that in the middle panel, the transition between the squeezed and standard tree-level results for $\ell=2$ and $\ell=4$ exhibits a much sharper discontinuity compared to the case for $\ell=0$. This may suggest that our choice of $f_\text{sq} = 0.5$ may have to be revisited in more careful investigations of the angular dependence of the covariance using our stitched results. We note also that the higher multipoles of the covariance $(\ell > 0)$ depend to a much higher degree on the anisotropic response coefficients $R_O(k)$, for which we currently only have extrapolations based on the physical reasoning described in Ref.~\cite{paper1}. 

The higher multipoles of the covariance matrix are much less studied in the literature than the monopole, since most investigations only consider the covariance of the angle-averaged matter power spectrum. One exception is Ref.~\cite{joachimpen2011}, who measure the angular dependence of the covariance using N-body simulations (see also Ref.~\cite{1999ApJ...527....1S} for an earlier perturbation theory calculation). Here, we do not attempt to perform detailed comparisons against these simulation results, given the lack of simulation-calibrated anisotropic response coefficients for their cosmology, and defer that to future work. Nevertheless, we point out that the prediction depicted in the left panel of Fig.~\ref{fig:angle} is in agreement with the hierarchy of the diagonal covariance elements shown in Fig.~10 of Ref.~\cite{joachimpen2011}.

\section{Summary and Discussion}\label{sec:summ}

We have described a calculation of the matter power spectrum covariance $\cov(\vk_1, \vk_2)$ based on perturbation theory augmented with specific resummed interaction vertices (the responses), which is applicable in all regimes of structure formation. More specifically, we describe the non-Gaussian part of the matter power spectrum covariance which is equivalent to the parallelogram configuration of the matter trispectrum, $T_m(\vk_1, -\vk_1, \vk_2, -\vk_2)$ (cf.~Eq.~(\ref{eq:covdef})). There are two other important contributions to the total matter covariance, namely the Gaussian diagonal term and the (also non-Gaussian) super-sample contribution, but these are both well understood. Our calculation is built upon the work of Ref.~\cite{paper1}, in which the authors have illustrated how the calculation of certain mode-coupling interactions in perturbation theory can be made accurate beyond the perturbative regime with the aid of power spectrum responses.

The $n$-th order power spectrum responses $\R_n$ describe the coupling of $n$ long-wavelength modes with the local nonlinear matter power spectrum (cf.~Eq.~(\ref{eq:Rndef})), and these responses can be measured accurately with separate universe simulations. The crucial and novel steps of our calculation consist essentially in the identification of the mode-coupling terms in the non-Gaussian covariance that can be described as power spectrum responses (or more technically, that can be resummed to all orders in perturbation theory using responses), thereby enabling efficient and accurate evaluation of these terms in kinematical regimes in which standard perturbation theory breaks down. The well-defined angular structure of the $\R_n$ also permits us to straightforwardly determine the angular dependence of the covariance matrix (cf.~Sec.~\ref{sec:angles}). Although the formalism presented here still needs as ingredients response measurements from simulations, we stress that the number of simulations that need to be performed for these measurements are {\it orders of magnitude} fewer than those that are needed for the direct, fully simulation-based estimation of the power spectrum covariance.

In this paper, we have worked explicitly at tree- and 1-loop-levels in the standard perturbation theory expansion of the non-Gaussian covariance (cf.~Eq.~(\ref{eq:covngexp})). At tree level (cf.~Sec.~\ref{sec:NGcovtree}), we have presented a way to stitch together response-based terms with terms from standard perturbation theory. At the 1-loop level (cf.~Sec.~\ref{sec:NGcov1loop}), we have seen that a response approach is particularly useful because a significant part of the contribution comes from the coupling of soft loop to hard external momenta, $p \ll k_1, k_2$, which are precisely interactions that power spectrum responses are able to capture. We have also pointed out, however, that our response-based 1-loop calculation still leaves important contributions to the covariance uncovered, but which can be added after some additional development (cf.~Secs.~\ref{sec:defnr} and \ref{sec:defhl}).

\begin{figure}[t!]
	\centering
	\includegraphics[width=\textwidth]{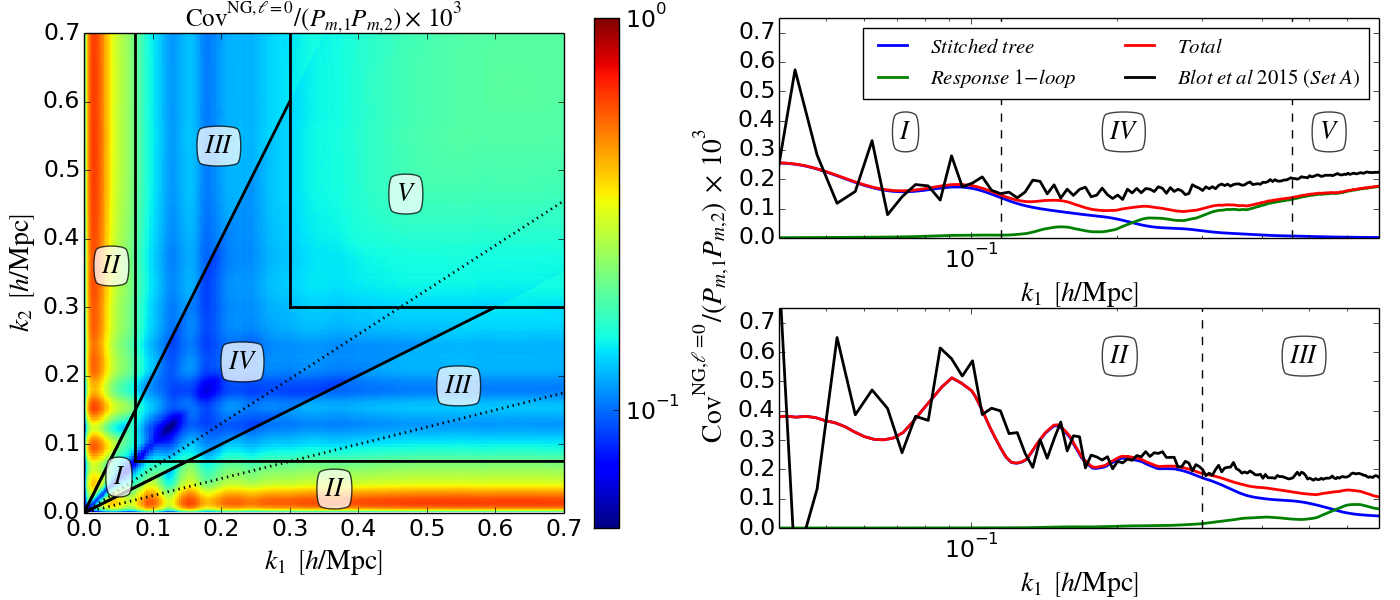}
	\caption{Summary of the various kinematic regimes of the structure of the matter covariance matrix. The color plot shows the non-Gaussian covariance (angle-averaged and at $z=0$) obtained by summing the stitched tree and response 1-loop results, as described in this paper. The regions bounded by the solid lines cover roughly the five kinematic regimes discussed in the text, as labeled (cf.~Sec.~\ref{sec:summ}). The right panels correspond to the two slices of constant $k_2/k_1 = 0.65$ (upper right) and $k_2/k_1 = 0.25$ (lower right) depicted by the dotted lines in the color plot, with the regimes identified as well, as labeled. As discussed in the text, the covariance matrix shown correctly describes regimes \emph{I} and \emph{II}. It also accounts for a majority of the contribution in the other regimes (taking the simulation results of Ref.~\cite{blot2015} to guide the eye). However, there are still important known contributions that can be added to further increase the accuracy of the calculation in regimes \emph{III}, \emph{IV} and \emph{V}.}
\label{fig:reg}
\end{figure}

In order to organize the discussion about which parts of the covariance are already captured by our description, and which require additional work, we can divide the parameter space of $\cov^{\rm NG, \ell=0}(k_1,k_2)$ into 5 kinematic regimes, as illustrated in Figure \ref{fig:reg}. This division naturally arises when distinguishing three regimes of wavenumber for $k_1$ and $k_2$: the linear regime ($k_i \ll \knl$), the quasilinear regime $k_i \lesssim \knl$, and the fully nonlinear regime $k \gtrsim \knl$. The right panels in \reffig{reg} show $\cov^{\rm NG, \ell=0}(k_1,k_2)$ as a function of $k_1$, while keeping $k_2/k_1$ constant, i.e., fixed level of \emph{squeezing}. We now briefly discuss each of these regimes, denoting as throughout $k_\text{soft} \equiv \min\{k_1,k_2\}$ and $k_\text{hard} \equiv \max\{k_1,k_2\}$.

\bigskip 

\hspace{0.2 cm} $\bullet$ \underline{\emph{I}: $k_1 \ll \knl,\  k_2 \ll \knl$.} In this regime, with both modes in the linear regime, the covariance is captured completely by the standard tree-level result (cf.~Eq.~(\ref{eq:covtree})).

\hspace{0.2 cm} $\bullet$ \underline{\emph{II}: $k_\text{soft} \ll \knl,\  k_\text{soft} \lesssim f_\text{sq} k_\text{hard}$, for any $k_\text{hard}$.} In this squeezed regime, with the soft mode being in the linear regime, $\cov^{\rm NG, \ell=0}(k_1, k_2)$ is exactly captured by the second-order power-spectrum response $\R_2$ (cf.~Eqs.~(\ref{eq:resptreecov})), up to corrections that scale as $\O((k_\text{soft}/k_\text{hard})^2)$, $\O((k_\text{soft}/\knl)^2)$.  Note that the result remains valid for any value of the hard momenta, including $k_\text{hard} > \knl$.

\hspace{0.2 cm} $\bullet$ \underline{\emph{III}: $k_\text{soft} \lesssim \knl,\  k_\text{soft} \lesssim f_\text{sq} k_\text{hard}$, for any $k_\text{hard}$.} This regime is still squeezed, but with quasilinear values of $k_\text{soft}$. The response tree-level result that fully determines regime \emph{II} still contributes, but now loop terms are no longer negligible and become increasingly important with increasing $k_\text{soft}$. While in this paper, we have included only the single response-type contribution that is present for generic configurations (cf.~\refeq{Cov1loop}), we expect that all loop contributions in the squeezed regime can be captured by responses, in the sense that non-response-type contributions are suppressed by $(k_\text{soft}/k_\text{hard})^2$. 

\hspace{0.2 cm} $\bullet$ \underline{\emph{IV}: $k_\text{soft} \sim k_\text{hard} \lesssim \knl$.} In this non-squeezed, quasi-linear regime, the standard tree level term still contributes non-negligibly and  the 1-loop contributions are important. In this regime, we expect the non-response-type 1-loop terms to be relevant as well. Higher-order loop terms also become increasingly relevant as $k_1$ and $k_2$ approach $\knl$.

\hspace{0.2 cm} $\bullet$ \underline{\emph{V}: $k_\text{soft} \sim k_\text{hard} \gtrsim \knl$.} In this regime, the tree-level and non-response-type 1-loop contributions are negligible and the result is dominated by response-type loop terms. In this paper, we have worked explicitly at 1-loop order, but higher-loop terms (cf.~Eq.~(\ref{eq:Cov2loop}) for two-loop) are expected to be significant as well. Crucially, we note that if higher loop contributions are not progressively suppressed, then the approach presented here will not be predictive in this regime.

\bigskip

The discussion points above motivate two immediate steps that can be taken to improve our prediction for $\cov^{\rm NG}$. One is the inclusion of 1-loop terms that cannot be described with power spectrum responses and which are expected to be important in regime \emph{IV}. These terms have already been derived and calculated in Ref.~\cite{2016PhRvD..93l3505B}. One can therefore include them into our model by following a ``stitching'' recipe similar to that employed at tree level in this paper. The other improvement is the inclusion of higher-loop response-type contributions, which is expected to result in relevant contributions to regimes \emph{III, IV,V}. This calculation is also crucial to establish the theoretical consistency of the approach presented here: our prediction on fully nonlinear scales is only robust if the higher-loop contributions can be shown to be progressively suppressed compared to the leading 1-loop contribution derived here. This could happen if the relevant angle-averages of higher-order responses are suppressed.

In regimes \emph{I-II}, on the other hand, the current calculation captures already the total leading contribution to the covariance. An interesting consequence of this is that comparisons to our calculation in this regime can therefore serve as useful validation checks of simulation-based estimates of the covariance.


\bigskip

In summary, we have paved the way towards the development of a physically motivated framework that enables an efficient calculation of the matter power spectrum covariance without any adjustable free parameters, and which is valid deeply into the nonlinear regime of structure formation where standard perturbative schemes break down. Our calculation can also be generalized to describe matter correlations at different redshift values (by making use of unequal-time power spectrum responses \cite{paper1}), which is useful for tomographic cosmic shear analyses. Compared to standard ways to estimate the covariance matrix with N-body simulations, an approach combining simulations with analytical results such as the one put forward here has the enormous advantage of requiring far less computational resources. A straightforward consequence of this is that robust and systematic studies of the dependence on cosmology of the covariance can be performed through this approach. The same can be said about the impact of baryonic effects on the covariance matrix. These constitute pieces of information that are very relevant for upcoming observational surveys such as Euclid \cite{2011arXiv1110.3193L} and LSST \cite{2012arXiv1211.0310L} 
whose statistical precision will be at a level sufficient to make these systematic effects on the covariance a pressing concern.

\begin{acknowledgments}

We thank Linda Blot for providing the numerical measurements of the power spectrum covariance, and Jean-Michel Alimi, Linda Blot, Pier-Stefano Corasaniti, Joachim Harnois-D\'eraps, Wayne Hu, Irshad Mohammed, Yann Rasera, Vincent Reverdy, Uro$\check{\rm s}$ Seljak and Zvonimir Vlah  for useful discussions. 

FS acknowledges support from the Marie Curie Career Integration Grant  (FP7-PEOPLE-2013-CIG) ``FundPhysicsAndLSS,'' and Starting Grant (ERC-2015-STG 678652) ``GrInflaGal'' from the European Research Council.

\end{acknowledgments}

\appendix 

\section{Feynman rules for cosmological perturbation theory}\label{app:feynman}

In this appendix, we list the Feynman rules we employ to calculate $n$-point functions in cosmological perturbation theory. Our conventions are based on the one used by Ref.~\cite{abolhasani/mirbabayi/pajer:2016}. The rules are as follows: 
\begin{enumerate}
\item An $n$-point correlation function is represented by a set of diagrams with $n$ outgoing external legs. 
\item Interaction vertices have $m\geq 2$ ingoing lines $\vp_1,\cdots,\vp_m$ coupling to a single outgoing line $\vp$, and each such vertex is assigned a factor
  \be
   m! F_m(\vp_1, \cdots, \vp_m) (2\pi)^3 \d_D(\vp-\vp_{1\cdots m})\,.
  \ee
We assign a negative (positive) sign to ingoing (outgoing) momenta. Each ingoing line has to be directly connected to a propagator (linear power spectrum).

\item Propagators are represented in our notation as vertices with 2 outgoing lines of equal momentum $k$ as {\includegraphicsbox[scale=0.8]{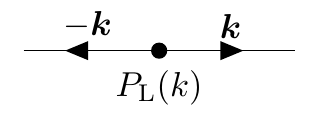}}, and they are assigned a factor $\Plin(k)$. To ease the notation, we often skip labeling these two outgoing lines (which line is which can always be inferred from momentum conservation).

\item All momenta that are not fixed in terms of momentum constraints are called loop momenta and are integrated over as
  \be
  \int_{\vp} \equiv \int \frac{d^3\vp}{(2\pi)^3}\,.
  \ee
A tree-level diagram is a diagram without any such loop integrals.
\item Each diagram is multiplied by the {\it symmetry factor}, which accounts for degenerate configurations of the diagram, as well as all nonequivalent labellings of external lines.
\end{enumerate}

\noindent To include response-type interactions, we consider one additional rule:

\begin{enumerate}\setcounter{enumi}{5}
\item Response-type interaction vertices have 2 (instead of 1) outgoing lines with momenta $\vk,\vk'$, and $n\geq 1$ incoming lines with momenta $\vp_a$. These vertices are only predictive in the limit where $\sum_a p_a \ll \min\{k,\knl\}$, but no restriction is placed on the magnitude of the outgoing momenta, which can be in the nonlinear regime. In our notation, we represent them as dashed blobs. Each such vertex is assigned a factor (cf.~Eq.~(\ref{eq:Rndef}))
  \be
 \frac12 \R_n(k; \cdots)\, P_m(k) (2\pi)^3 \d_D(\vk+\vk' - \vp_{1\cdots a})\,,
  \ee
where the dots in the argument of $\R_n$ denote all the relevant cosine angles and soft momenta magnitude ratios that exist at a given order $n$ (described in detail in Sec.~2 of Ref.~\cite{paper1}). The factor $1/2$ serves to cancel the trivial permutation $\vk \leftrightarrow \vk'$, which is always present in the response vertex.
\end{enumerate}

\section{Response expressions}\label{app:ro}

In this appendix, we show the decomposition of the second-order response $\R_2$ as derived in Ref.~\cite{paper1}, including the expression of the response coefficients that we use in our covariance calculations.

For general kinematic configurations $\mu_1$, $\mu_2$, $\mu_{12}$ and $f_{12} = p_1/p_2$, the second-order matter power spectrum response is given by (in the Eulerian decomposition; see also Ref.~\cite{paper1} for the Lagrangian decomposition and Ref.~\cite{bertolini1} for another example of the angular decomposition)

\bq\label{eq:PR2_angle}
&&\R_2(k; \mu_1,\mu_2,\mu_{12}, f_{12}) =  R_1(k)  \Bigg[{\frac{5}{7}} +  \frac{\mu_{12}}{2}\big(f_{12} + \frac{1}{f_{12}}\big)+ \frac27 \mu_{12}^2 \Bigg] \nonumber \\
&&\hspace*{1cm}+ R_K(k)\Bigg[\mu_1 \mu_2 \mu_{12} - \frac13 {\mu_{12}^2 } + \frac57 \left(\frac{( \mu_1 + f_{12} \mu_2)^2}{1 + f_{12}^2 + 2 f_{12} \mu_{12}} - \frac13 \right) (1 - \mu_{12}^2)  \nonumber \\ 
&& \hspace*{1cm}\qquad\qquad + \frac12  \mu_{12} \Bigg( \left(\mu_1^2 - \frac13\right)f_{12} + \left(\mu_2^2 - \frac13\right)\frac{1}{f_{12}} \Bigg) \Bigg] \nonumber \\
&&\hspace*{1cm}+ \frac12 R_2(k) + \frac12 R_{K \d}(k) \Bigg[\mu_1^2 + \mu_2^2 - \frac23 \Bigg] + R_{K^2}(k) \Bigg[\mu_{12}^2 - \frac13\Bigg]  \nonumber \\
&&\hspace*{1cm}+ R_{K.K}(k) \Bigg[\mu_1 \mu_2 \mu_{12} {-\frac13\mu_1^2-\frac13\mu_2^2+\frac19}\Bigg] + R_{KK}(k) \Bigg[\mu_1^2\mu_2^2 - \frac13\left(\mu_1^2 + \mu_2^2\right) + \frac19\Bigg] \nonumber \\ 
&&\hspace*{1cm}+  \frac32 R_{\Ote}(k) \left(\frac{( \mu_1 + f_{12} \mu_2)^2}{1 + f_{12}^2 + 2 f_{12} \mu_{12}} - \frac13 \right) (1 - \mu_{12}^2)\,.
\eq
Setting $\mu_1 = \mu$, $\mu_2 = -\mu$, $\mu_{12}=-1$ and $f_{12} = 1$ yields Eq.~(\ref{eq:PR2_anglecov}), which is the configuration of $\R_2$ which enters the covariance calculations. The nonlinear extrapolation recipe of the $R_O(k)$ presented in Ref.~\cite{paper1} results in the following expressions:
\bq\label{eq:rsnl}
R_1(k) &=& \mbox{Measurement from Ref.~\cite{response}},\nonumber \\
R_K(k) &=& \frac{12}{13} G_1^{\rm nl}(k) - \fkonenl, \nonumber \\
R_2(k) &=& \mbox{Measurement from Ref.~\cite{response}}, \nonumber \\
R_{K\delta}(k) &=& \frac{1518}{1813}\left[\frac{8}{21}G_1^{\rm nl}(k) + G_2^{\rm nl}(k)\right] + \frac{41}{22}\left[-\frac29 - \frac23G_1^{\rm nl}(k)\right]\fkonenl + \frac13\fktwonl, \nonumber \\
R_{K^2}(k) &=& \frac{1}{21}G_1^{\rm nl}(k) - \frac{1}{6}\fkonenl,\nonumber \\ 
R_{K.K}(k) &=& -\frac{22}{13}G_1^{\rm nl}(k) +\frac32 \fkonenl, \nonumber \\
R_{KK}(k) &=& \frac{1476}{1813}\left[\frac{8}{21}G_1^{\rm nl}(k) + G_2^{\rm nl}(k)\right] + \frac{69}{44}\left[-\frac29 - \frac23G_1^{\rm nl}(k)\right]\fkonenl + \frac12\fktwonl, \nonumber \\
R_{\Ote}(k) &=& -\frac{92}{273}G_1^{\rm nl}(k) +\frac13\fkonenl\,,
\eq
where $G_n^{\rm nl}(k)$ correspond to the so called {\it growth-only} response functions measured in Ref.~\cite{response}. The nonlinear shapes of $R_1(k)$ and $R_2(k)$ are also those measured directly from the separate universe simulations of Ref.~\cite{response}. Reference \cite{response} found that the different isotropic response coefficients ($R_1$, $R_2$ and $R_3$) are similar in shape. In the context of the halo model, this can be explained by the characteristic transition between the 2-halo and 1-halo regimes, which leads to a peak of the response coefficients around $k\sim 0.5\kunit$, with a suppression at higher $k$. The derivation of the above expressions for the $R_O(k)$ assumes that these halo model-based considerations extend also to the case of the anisotropic response coefficients.

In the exercise displayed in Fig.~\ref{fig:fsq} below, we use the tree-level expressions of the $R_O(k)$. Starting from Eq.~(\ref{eq:rsnl}), these can be obtained with the substitutions (see Refs.~\cite{response, paper1} for more details) $P_m(k) \rightarrow \Plin(k)$, $G_1^{\rm nl} \rightarrow G_1^{\rm tree} = 26/21$, $G_2^{\rm nl} \rightarrow G_2^{\rm tree} = 3002/1323$ and
\bq\label{eq:r1r2tree}
R_1(k) &\rightarrow& 1 - \frac13 \frac{d\ln \Plin(k)}{d\ln k} + G^{\rm tree}_1, \nonumber \\
R_2(k) &\rightarrow& \left(\frac{8}{21} G_1^{\rm tree} + G_2^{\rm tree}\right )+ \left(-\frac{2}{9} - \frac23 G_1^{\rm tree}\right)\fkone + \frac19\fktwo.
\eq

\section{Multipoles of the response tree-level and 1-loop covariance}\label{app:analy}

In this appendix, we collect all the Legendre multipoles of the tree-level $\cov_{\R\text{-tree}}^{\rm NG, \ell}(k_1, k_2)$ and 1-loop level $\cov_{\R\text{-1loop}}^{\rm NG, \ell}(k_1, k_2)$ covariance in the response approach, which can be given analytically.

At tree-level, using Eqs.~(\ref{eq:PR2_anglecov}), (\ref{eq:resptreecov}) and (\ref{eq:angleaver}), we can write all non-vanishing multipoles ($\ell=0,2,4$) as
\bq
\label{eq:multipoles_sqtree0} V \cov_{\R\text{-tree}}^{\rm NG, \ell=0}(k_1, k_2) &=& 2\bigg[\A(k_\text{hard})+\frac15\C(k_\text{hard})\bigg]\Plin(k_\text{soft})^2P_m(k_\text{hard}), \\
\label{eq:multipoles_sqtree2} V \cov_{\R\text{-tree}}^{\rm NG, \ell=2}(k_1, k_2) &=& 2\bigg[\B(k_\text{hard})+\frac27\C(k_\text{hard})\bigg]\Plin(k_\text{soft})^2P_m(k_\text{hard}), \\
\label{eq:multipoles_sqtree4} V \cov_{\R\text{-tree}}^{\rm NG, \ell=4}(k_1, k_2) &=& 2\bigg[\frac{18}{35}\C(k_\text{hard})\bigg]\Plin(k_\text{soft})^2P_m(k_\text{hard})\,,
\eq
where, as in the main text, $k_\text{soft} = {\rm min}\{k_1, k_2\}$ and $k_\text{hard} = {\rm max}\{k_1, k_2\}$.

Similarly, but for the 1-loop expressions, we can combine Eqs.~(\ref{eq:PR2_anglecov}), (\ref{eq:cov1loop_res}) and (\ref{eq:angleaver}) to find
\bq
\label{eq:cov1loop_l0} V \cov_{\R\text{-1loop}}^{\rm NG, \ell=0}(k_1, k_2) &=& \frac{2P_m(k_1)P_m(k_2)}{(2\pi)^2}\Bigg[\int_0^{ p_\text{max}}p^2[\Plin(p)]^2 {\rm d}p\Bigg] \nonumber \\
&& \ \ \ \ \ \ \ \ \ \times \Bigg[2\A_1\A_2 + \frac25\big(\A_1\C_2 + \A_2\C_1\big) + \frac{2}{25}\C_1\C_2\Bigg] \\
\label{eq:cov1loop_l2}V \cov_{\R\text{-1loop}}^{\rm NG, \ell=2}(k_1, k_2) &=& \frac{2P_m(k_1)P_m(k_2)}{(2\pi)^2}\Bigg[\int_0^{ p_\text{max}}p^2[\Plin(p)]^2 {\rm d}p\Bigg] \nonumber \\ 
&& \ \ \ \ \ \ \ \ \ \times \Bigg[\frac25\B_1\B_2 + \frac{4}{35}\big(\B_1\C_2 + \B_2\C_1\big) + \frac{8}{245}\C_1\C_2\Bigg] \\
\label{eq:cov1loop_l4} V\cov_{\R\text{-1loop}}^{\rm NG, \ell=4}(k_1, k_2) &=& \frac{2P_m(k_1)P_m(k_2)}{(2\pi)^2}\Bigg[\int_0^{ p_\text{max}}p^2[\Plin(p)]^2 {\rm d}p\Bigg] \Bigg[\frac{72}{1225}\C_1\C_2\Bigg],
\eq
while all other multipoles vanish exactly (here, $\A_i \equiv \A(k_i)\ (i=1,2)$, and similarly for $\B_i$ and $\C_i$).

\section{Determination of $f_\text{sq}$}\label{app:fsq}

In this appendix, we describe the procedure used to choose the value of $f_\text{sq}$, which controls the transition from the squeezed to the non-squeezed branches in our stitched tree-level result of Eq.~(\ref{eq:stitched}).

\begin{figure}
	\centering
	\includegraphics[width=\textwidth]{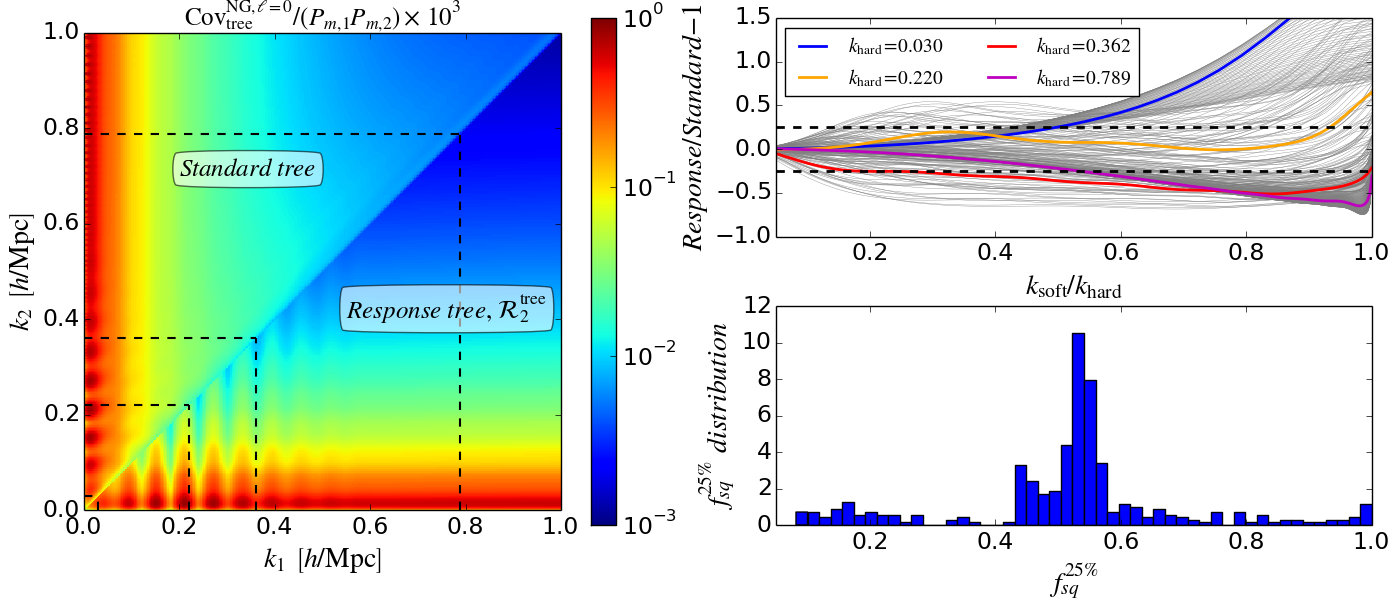}
	\caption{Determination of the value of $f_\text{sq}$ that controls the transition from squeezed to non-squeezed expressions. The color plot on the left shows the non-Gaussian angle-averaged tree-level covariance ($\ell=0$ in Eq.~(\ref{eq:angleaver})). The upper triangle shows the standard tree-level result (upper branch in Eq.~(\ref{eq:stitched})), while the lower panel shows the response tree-level result (lower branch in Eq.~(\ref{eq:stitched})), obtained with the tree-level expression of $\R_2$ (cf.~Appendix~\ref{app:ro}). The colored lines in the upper right panel show the difference of the covariance along the vertical dashed lines on the color plot, relative to the covariance along the corresponding horizontal dashed lines, plotted as a function of $k_\text{soft}/k_\text{hard}$. The grey curves show the same for all other pairs of slices of the covariance matrix. The lower right panel shows the distribution of the values of $f_\text{sq}^{25\%} = k_\text{soft}/k_\text{hard}$ at which the relative differences drop below $25\%$. The results correspond to $V = 656.25\ h^{-3} {\rm Mpc}^3$ and $P_{m,i}\equiv P_m(k_i)$ in the title of the color plot.}
\label{fig:fsq}
\end{figure}

The upper triangle of the color plot in Fig.~\ref{fig:fsq} shows the angle-averaged standard tree-level covariance $\cov_\text{SPT-tree}^{\rm NG, \ell=0}(k_1, k_2)$  ($\ell = 0$ in Eq.~(\ref{eq:angleaver})) using the upper branch of Eq.~(\ref{eq:stitched}). The lower triangle shows the response tree-level result $\cov_{\R\text{-tree}}^{\rm NG, \ell=0}(k_1, k_2)$, but using the tree-level limit of the second-order response $\R_2^{\rm tree}$ (cf.~Appendix \ref{app:ro}). By construction, these two results are the same in the limit in which one of the modes is much smaller than the other. A reasonable choice of $f_\text{sq}$ should therefore be one that ensures a sufficiently smooth transition between the two branches of Eq.~(\ref{eq:stitched}) at $k_\text{soft} = f_\text{sq} k_\text{hard}$. The upper right panel of Fig.~\ref{fig:fsq} shows the relative difference between the two non-Gaussian tree level expressions, plotted as a function of $k_\text{soft}/k_\text{hard}$. The several lines shown (grey) correspond to different values of $k_\text{hard}$. To guide the eye, the colored curves show the relative difference between the two covariance results along the dashed lines overlaid in the color plot. We consider the two results to be sufficiently close to one another when the relative difference drops below the $25\%$ mark (dashed lines in the upper right panel). The distribution of the values of $f_\text{sq}^{25\%} = k_\text{soft}/k_\text{hard}$ when this happens is shown in the lower right panel of Fig.~\ref{fig:fsq}. The mean and median of the distribution are both $\approx 0.5$, and for that reason we choose $f_\text{sq} = 0.5$. The choice of the $25\%$ figure is somewhat arbitrary, but small variations around it do not affect our tree level results significantly.


\section{Explicit relation between the product of two $\R_2$ responses and 1-loop covariance terms}\label{app:derivation}

In this appendix, we show explicitly which of the 1-loop covariance terms are captured by the response diagram of Eq.~(\ref{eq:Cov1loop}). In order to compare with the 1-loop expressions in standard perturbation theory, we have to insert the tree-level expression for the response. We then obtain 
\bq\label{eq:Cov1loop_tree}
&& {\rm Eq.~(\ref{eq:Cov1loop})}_{\text{tree-level\ }\R_2} : \nonumber \\ 
& = & \int \frac{{\rm d}^3p}{(2\pi)^3} 2 \R_2^{\rm tree}(k_1,\mu_1,-\mu_1,-1,1)  \R_2^{\rm tree}(k_2,\mu_2,-\mu_2,-1,1) \Plin(k_1)\Plin(k_2)[\Plin(p)]^2 \nonumber \\
 &=& \int \frac{{\rm d}^3p}{(2\pi)^3} \Bigg[72 \fiii(\vk_1, \vp, -\vp)\fiii(\vk_2, \vp, -\vp) \nonumber \\ 
&+& 48\Big\{\fiii(\vk_1, \vp, -\vp)[\fii(\vp-\vk_2, -\vp)]^2 + (\vk_1 \leftrightarrow \vk_2)\Big\} \nonumber \\
&+& 8\Big\{[\fii(\vp+\vk_1, -\vp)]^2+[\fii(\vp-\vk_1, -\vp)]^2\Big\}\Big\{[\fii(\vp+\vk_2, -\vp)]^2+[\fii(\vp-\vk_2, -\vp)]^2\Big\} \Bigg] \nonumber \\ 
&& \ \ \ \ \ \ \ \ \ \ \ \ \ \ \ \ \ \ \ \ \ \ \ \ \ \ \ \ \ \ \ \ \ \ \ \ \ \ \ \ \ \ \ \ \ \ \ \ \ \ \ \ \ \ \ \ \ \ \ \ \ \ \ \ \ \ \ \ \times \Plin(k_1)\Plin(k_2)[\Plin(p)]^2 \nonumber \\
\eq
where we have used 
\bq\label{eq:R2tree}
\R_2^{\rm tree}(k,\mu,-\mu,-1,1) = 6\fiii(\vk, \vp, -\vp) + 2[\fii(\vp+\vk, -\vp)]^2 + 2[\fii(\vp-\vk, -\vp)]^2\,,
\eq
which holds in the limit $p \ll k$ as derived in Ref.~\cite{paper1} using the squeezed-limit tree-level trispectrum.\footnote{
From the considerations in Sec.~4.2 of Ref.~\cite{paper1}, we can write
\bq
&& 2\R_2^{\rm tree}(k, \mu_1, \mu_2, \mu_{12}, p_1/p_2)\Plin(k')\Plin(p_1)\Plin(p_2) =  \lim_{p_1, p_2 \to 0}\Big[6\fiii(\vk, \vp_1, \vp_2)\Plin(k)\Plin(p_1)\Plin(p_2)  \nonumber \\
&& \ \ \ \ \ \ \ \ \ \ \ \ \ \ \ \ \ \ \ \ \ \ \ \ \ \ \ \ + 4\fii(-\vp_1, \vk+\vp_1)\fii(\vp_2, \vk+\vp_1)\Plin(k+p_1)\Plin(p_1)\Plin(p_2) +  (\vk \leftrightarrow \vk')\Big],
\eq
where as everywhere in this paper, the limit means retaining terms at leading order in $p_1$ and $p_2$, and $\vk' = -\vk - \vp_1 - \vp_2$. Specifying to the covariance configuration, $\vp_1 = \vp$, $\vp_2 = -\vp$ and approximating $\Plin(k \pm p) \approx \Plin(k)$, we arrive at Eq.~(\ref{eq:R2tree}).
}
Our goal is to demonstrate that Eq.~(\ref{eq:Cov1loop_tree}) is indeed obtained by summing the 1-loop trispectrum terms that are $\O([P(p)]^2)$ in the covariance configuration. This demonstration can be carried out fairly straightforwardly, despite involving tedious counting of momentum permutations. For completeness, we shall nevertheless lay down a few intermediate steps of the derivation below. 

The matter trispectrum that contributes to the covariance at 1-loop (which is directly related to the covariance via a volume factor), $T_m^{\oneloop}(\vk_1, -\vk_1, \vk_2, -\vk_2) \equiv \bar{T}_m^{\oneloop}(\vk_1, \vk_2)$, has contributions from nine types of diagrams (we do not list all of them for brevity, but see e.g.~Appendix A of Ref.~\cite{2016PhRvD..93l3505B}). Out of these nine diagrams, there are three which contain permutations of momenta that give rise to terms that are quadratic in the power spectrum of the loop momentum $[P(p)]^2$, with the remaining six diagrams being linear in $P(p)$ (recall the discussion around Eqs.~(\ref{eq:Pp}) and (\ref{eq:Pp2})):
\bq\label{eq:oneloopsplit}
\bar{T}_m^{\oneloop}(\vk_1, \vk_2) = \bar{T}^{\oneloop}_{m, [P(p)]^2}(\vk_1, \vk_2) + \bar{T}^{\oneloop}_{m, P(p)}(\vk_1, \vk_2),
\eq
in which the split on the right-hand side separates the diagrams by their order in $P(p)$, as indicated in the subscripts. From hereon in this appendix, we also drop the subscript $_m$ to ease the notation. The response-based definition of Eq.~(\ref{eq:Cov1loop_tree}) describes only the terms that contribute to $T^{\oneloop}_{[P(p)]^2}(\vk_1, -\vk_2, \vk_2, -\vk_2)$, which are
\bq\label{eq:Pq2terms}
\bar{T}^{\oneloop}_{[P(p)]^2}(\vk_1, \vk_2) = \bar{T}_{2222}(\vk_1, \vk_2) + \bar{T}_{3311b}(\vk_1, \vk_2) + \bar{T}_{3221c}(\vk_1, \vk_2),
\eq
where we have adopted the same notation as in Ref.~\cite{2016PhRvD..93l3505B}. Next, we write these three diagrams explicitly and demonstrate that their sum yields exactly Eq.~(\ref{eq:Cov1loop_tree}).

We start with $T_{2222}$, which consists of three terms given by
\bq\label{eq:t2222}
T_{2222}(\vk_1, \vk_2, \vk_3, \vk_4) &=& 16 \int \frac{{\rm d}^3p}{(2\pi)^3} \fii(\vp, -\vk_1 - \vp)\fii(-\vp, \vp - \vk_2)\fii(\vk_2-\vp, \vp - \vk_{23}) \nonumber \\
&&  \times \fii(\vk_{23} - \vp, \vp + \vk_1) \Plin(p)\Plin(|\vp-\vk_2|)\Plin(|\vp-\vk_{23}|)\Plin(|\vp+\vk_1|) \nonumber \\ 
&& + (\vk_1 \leftrightarrow \vk_2) + (\vk_2 \leftrightarrow \vk_3).
\eq
Note that out of all possible 24 permutations of four elements, only three correspond to distinct diagrams. In the covariance configuration ($\vk_2=-\vk_1$, $\vk_4 = -\vk_3$), only two of the permutations give contributions $\propto [\Plin(p)]^2$ for general $\vk_1, \vk_3$.\footnote{For instance, in the covariance configuration, the spectra factors of the permutation written explicitly in Eq.~(\ref{eq:t2222}) become $\Plin(p)[\Plin(|\vp+\vk_1|)]^2\Plin(|\vp+\vk_1-\vk_3|)$. Noting that we can transform $\vp+\vk_1 \to \vp$ because we are integrating over $\vp$, we can write $[\Plin(p)]^2\Plin(|\vp-\vk_1|)\Plin(|\vp-\vk_3|)$.} The other permutation can also be $\propto [\Plin(p)]^2$, but only if $\vk_1 = \vk_3$, i.e., if the two wavevectors have the same magnitude and are aligned. After angle-averaging, this term contributes negligibly, so we do not consider it further. We note however that this term can be modeled accurately with $\R_2$ as well by modifying Eq.~(\ref{eq:Cov1loop}) to include a term proportional to $\delta_D(\vk_1 - \vk_3)$.  One can then straightforwardly work out that (relabeling $\vk_3 \rightarrow \vk_2$ and with the limit $p \ll k_1, k_2$ understood)
\bq\label{eq:t2222cov}
&& \frac{\bar{T}_{2222}(\vk_1, \vk_2)}{\Plin(k_1)\Plin(k_2)}\bigg\rvert_{\R_2} = 16 \int \frac{{\rm d}^3p}{(2\pi)^3} \bigg[[\fii(-\vp, \vp+\vk_1)]^2[\fii(-\vp, \vp+\vk_2)]^2 \nonumber \\ 
&& \ \ \ \ \ \ \ \ \ \ \ \ \ \ \ \ \ \ \ \ \ \ \ \ \ \ \ \ \ \ \ \ \ \ \ \ \ \ \ \ \ \ \ \ \ \ \ \ \ \ \ \ +[\fii(-\vp, \vp+\vk_1)]^2[\fii(-\vp, \vp-\vk_2)]^2\bigg]  [\Plin(p)]^2 \nonumber \\
&=& 8 \int \frac{{\rm d}^3p}{(2\pi)^3} \bigg[[\fii(-\vp, \vp+\vk_1)]^2 + [\fii(-\vp, \vp-\vk_1)]^2\bigg] \bigg[[\fii(-\vp, \vp+\vk_2)]^2 + [\fii(-\vp, \vp-\vk_2)]^2\bigg] \nonumber \\
&& \ \ \ \ \ \ \ \ \ \ \ \ \ \ \times [\Plin(p)]^2,
\eq
which matches exactly the contribution in Eq.~(\ref{eq:Cov1loop_tree}) that is proportional to the product of four $\fii$ kernels. The second line is obtained from the first by using the fact that one can transform the loop momentum as $\vp \rightarrow -\vp$ in the integrand. The subscript $\R_2$ indicates that we consider only the mode permutations that are $\propto [\Plin(p)]^2$ for general $\vk_1, \vk_2$.

Similarly, the $T_{3311b}$ term can be written as
\bq\label{eq:t3311b}
T_{3311b}(\vk_1, \vk_2, \vk_3, \vk_4) &=& T_{3311b}^A(\vk_1, \vk_2, \vk_3, \vk_4) + T_{3311b}^B(\vk_1, \vk_2, \vk_3, \vk_4) + T_{3311b}^C(\vk_1, \vk_2, \vk_3, \vk_4), \nonumber \\
\eq
with
\bq\label{eq:t3311bA}
T_{3311b}^A(\vk_1, \vk_2, \vk_3, \vk_4) &=& \Bigg[18 \int \frac{{\rm d}^3p}{(2\pi)^3} \fiii(\vp, -\vp - \vk_{14}, \vk_4) \fiii(-\vp, \vp + \vk_{14}, \vk_3) \nonumber \\ 
&& \times\Plin(p)\Plin(|\vp+\vk_{14}|)\Plin(k_3)P(k_4) + (\vk_2 \leftrightarrow \vk_3)\Bigg]  \nonumber \\
&& \ \ \ \ \ \ \ \ \ \ \ \ \ \ \ \ \ \ \ \ \ \ \ \ \ \ \ \ \ \ \ \ + (\vk_1 \leftrightarrow \vk_4),
\eq
\bq\label{eq:t3311bB}
T_{3311b}^B(\vk_1, \vk_2, \vk_3, \vk_4) &=& \Bigg[18 \int \frac{{\rm d}^3p}{(2\pi)^3} \fiii(\vp, -\vp - \vk_{24}, \vk_4) \fiii(-\vp, \vp + \vk_{24}, \vk_3) \nonumber \\ 
&\times&\Plin(p)\Plin(|\vp+\vk_{24}|)\Plin(k_3)P(k_4) + (\vk_2 \leftrightarrow \vk_4)\Bigg]  \nonumber \\
&& \ \ \ \ \ \ \ \ \ \ \ \ \ \ \ \ \ \ \ \ \ \ \ \ \ \ \ \ \ \ \ \ + (\vk_1 \leftrightarrow \vk_3),
\eq
\bq\label{eq:t3311bC}
T_{3311b}^C(\vk_1, \vk_2, \vk_3, \vk_4) &=& \Bigg[18 \int \frac{{\rm d}^3p}{(2\pi)^3} \fiii(\vp, -\vp - \vk_{34}, \vk_4) \fiii(-\vp, \vp + \vk_{34}, \vk_2) \nonumber \\ 
&\times&\Plin(p)\Plin(|\vp+\vk_{34}|)\Plin(k_2)\Plin(k_4) + (\vk_1 \leftrightarrow \vk_2)\Bigg]  \nonumber \\
&& \ \ \ \ \ \ \ \ \ \ \ \ \ \ \ \ \ \ \ \ \ \ \ \ \ \ \ \ \ \ \ \ + (\vk_3 \leftrightarrow \vk_4),
\eq
where in total there are now twelve distinct permutations contributing to $T_{3311b}(\vk_1, \vk_2, \vk_3, \vk_4)$. Setting $\vk_2=-\vk_1$ and $\vk_4 = -\vk_3$, one finds that only the four permutations of $T_{3311b}^C$ are $\propto [\Plin(p)]^2$ for any $\vk_1$, $\vk_3$. Their contribution is obtained straightforwardly as (relabeling $\vk_3 \rightarrow \vk_2$)
\bq\label{eq:t3311bcov}
\frac{\bar{T}_{3311b}(\vk_1, \vk_2)}{\Plin(k_1)\Plin(k_2)}\bigg\rvert_{\R_2} &=& 72 \int \frac{{\rm d}^3p}{(2\pi)^3} \fiii(\vk_1, \vp, -\vp) \fiii(\vk_2, \vp, -\vp) [\Plin(p)]^2,
\eq
which matches the corresponding term in Eq.~(\ref{eq:Cov1loop_tree}). Here, similarly to the case of the $T_{2222}$ term above, we have also skipped writing the contribution from terms that are $\propto [\Plin(p)]^2$ when $\vk_1=\vk_2$ (the four permutations of $T_{3311b}^A$), as well as, $\vk_1=-\vk_2$ (the four permutations of $T_{3311b}^B$).

Finally, the $T_{3221c}$ term has also twelve terms, which we can write explicitly as
\bq\label{eq:t3221c}
T_{3221c}(\vk_1, \vk_2, \vk_3, \vk_4) &=& T_{3221c}^A(\vk_1, \vk_2, \vk_3, \vk_4) + T_{3221c}^B(\vk_1, \vk_2, \vk_3, \vk_4) + T_{3221c}^C(\vk_1, \vk_2, \vk_3, \vk_4), \nonumber \\
\eq
with
\bq\label{eq:t3221cA}
T_{3221c}^A(\vk_1, \vk_2, \vk_3, \vk_4) &=& \Bigg[24 \int \frac{{\rm d}^3p}{(2\pi)^3}\fiii(\vp, -\vp-\vk_{14}, \vk_4)\fii(-\vp, \vp-\vk_2)\fii(\vk_2-\vp, \vp+\vk_{14}) \nonumber \\ 
&& \ \ \ \ \ \ \ \ \ \ \ \ \ \ \ \ \ \times \Plin(p)\Plin(|\vp-\vk_2|)\Plin(|\vp+\vk_{14}|)\Plin(k_4) + (\vk_1 \leftrightarrow \vk_4) \Bigg] \nonumber \\ 
&+& (\vk_1 \leftrightarrow \vk_2),
\eq
\bq\label{eq:t3221cB}
T_{3221c}^B(\vk_1, \vk_2, \vk_3, \vk_4) &=& \Bigg[24 \int \frac{{\rm d}^3p}{(2\pi)^3}\fiii(\vp, -\vp-\vk_{12}, \vk_2)\fii(-\vp, \vp-\vk_3)\fii(\vk_3-\vp, \vp+\vk_{12}) \nonumber \\ 
&& \ \ \ \ \ \ \ \ \ \ \ \ \ \ \ \ \ \times \Plin(p)\Plin(|\vp-\vk_3|)\Plin(|\vp+\vk_{12}|)\Plin(k_2) + (\vk_1 \leftrightarrow \vk_2) \Bigg] \nonumber \\ 
&+& (\vk_1 \leftrightarrow \vk_3),
\eq
\bq\label{eq:t3221cC}
T_{3221c}^C(\vk_1, \vk_2, \vk_3, \vk_4) &=& \Bigg[24 \int \frac{{\rm d}^3p}{(2\pi)^3}\fiii(\vp, -\vp-\vk_{34}, \vk_4)\fii(-\vp, \vp-\vk_1)\fii(\vk_1-\vp, \vp+\vk_{34}) \nonumber \\ 
&& \ \ \ \ \ \ \ \ \ \ \ \ \ \ \ \ \ \times \Plin(p)\Plin(|\vp-\vk_1|)\Plin(|\vp+\vk_{34}|)\Plin(k_4) + (\vk_3 \leftrightarrow \vk_4) \Bigg] \nonumber \\ 
&+& (\vk_1 \leftrightarrow \vk_4).
\eq
After setting $\vk_2=-\vk_1$, $\vk_4 = -\vk_3$, we find four terms that are $\propto [\Plin(p)]^2$ (which come from $T_{3221c}^B$ and $T_{3221c}^C$) for general $\vk_1, \vk_3$, which we can write as (relabeling $\vk_3 \rightarrow \vk_2$ and interpreting $p \ll k_1, k_2$)
\bq\label{eq:t3221ccov}
\frac{\bar{T}_{3221c}(\vk_1, \vk_2)}{\Plin(k_1)\Plin(k_2)}\bigg\rvert_{\R_2} &=& 48 \int \frac{{\rm d}^3p}{(2\pi)^3} \Big[ \fiii(\vk_1, \vp, -\vp) [\fii(\vp, \vk_2 - \vp)]^2 + (\vk_1 \leftrightarrow \vk_2)\Big] [\Plin(p)]^2, \nonumber \\
\eq
which is exactly the remaining term in Eq.~(\ref{eq:Cov1loop_tree}).  Once again, we have skipped explicitly writing cases when $\vk_1 = \pm \vk_2$.

As discussed in the main body of the text, the generalization of Eq.~(\ref{eq:Cov1loop_tree}) to the nonlinear regime (that is, promoting $\Plin(k_1)$, $\Plin(k_2)$ and $\R_2^{\rm tree}$ to their nonlinear versions) then allows one to capture infinitely many higher-loop contributions via the simulation-calibrated nonlinear responses, while always considering only a single soft loop.

\section{Comparison to the covariance matrix model of Ref.~\cite{mohammed1}}\label{app:mohammed}

In this appendix, we compare our calculation of the covariance with the model put forward in Ref.~\cite{mohammed1}. In the latter, the total angle-averaged covariance is given by
\bq
\cov_{\mbox{\footnotesize{Ref.~\cite{mohammed1}}}}(k_1, k_2) = \cov^{\rm G}(k_1, k_2) + \cov_{\mbox{\footnotesize{Ref.~\cite{mohammed1}}}}^{\rm NG, tree}(k_1, k_2) + \cov_{\mbox{\footnotesize{Ref.~\cite{mohammed1}}}}^{\rm NG, 1-loop}(k_1, k_2). \nonumber \\
\eq
The term $\cov^{\rm G}(k_1, k_2)$ is the Gaussian result given by Eq.~(\ref{eq:gauss_mono}). The tree-level contribution $\cov_{\mbox{\footnotesize{Ref.~\cite{mohammed1}}}}^{\rm NG, tree}(k_1, k_2)$ is given by
\bq\label{eq:treemohammed}
\cov_{\mbox{\footnotesize{Ref.~\cite{mohammed1}}}}^{\rm NG, tree}(k_1, k_2) = \cov_\text{SPT-tree}^{\rm NG, \ell=0}(k_1, k_2) \frac{P_m(k_1)P_m(k_2)}{\Plin(k_1)\Plin(k_2)},
\eq
i.e., it is obtained by multiplying the standard perturbation theory tree-level result with a correction factor that is the ratio of nonlinear to linear spectra. Concretely, in the results shown in Ref.~\cite{mohammed1}, this correction is realized by comparing the tree-level covariance divided by the linear spectra with the simulation covariance estimates divided by the nonlinear spectra. While there is no obvious physical justification for this rescaling of the tree-level trispectrum, we will see below that it improves the model significantly when compared to simulation measurements.

The 1-loop term is obtained with the aid of a calculation based on functional derivatives of the small-scale power spectrum with respect to the large-scale one. Explicitly, the end result is given by (see Ref.~\cite{mohammed1} for details)
\bq\label{eq:1loopmohammed}
\cov_{\mbox{\footnotesize{Ref.~\cite{mohammed1}}}}^{\rm NG, 1-loop}(k_1, k_2) = \frac{1}{V\pi^2} \left(\frac{P_m(k_1)P_m(k_2)}{\Plin(k_1)\Plin(k_2)}\right) \int_0^{\infty} {\rm d}p \frac{p^2\left[\Plin(p)\right]^2}{\big(1+(p/p_\text{nl})^2\big)^2} V(p, k_1) V(p, k_2), \nonumber \\
\eq
with
\bq\label{eq:Vmohammed}
V(p, k) = 2\int {\rm d}\mu [\fii(\vk-\vp, \vp)]^2\Plin(|\vk-\vp|) + 3\Plin(k)\int {\rm d}\mu \fiii(\vk, \vp, -\vp).
\eq
In the limit $p/k \rightarrow 0$ we obtain
\bq\label{eq:Vsqmohammed}
V(p, k) &=& \left(\frac{2519}{2205} - \frac{47}{105}\fkone + \frac{1}{10}\fktwo\right)\Plin(k) \nonumber \\
&=& \frac{\Plin(k)}{2}\int {\rm d}\mu \R_2^{\rm tree}(k, \mu, -\mu, -1,1).
\eq
In this $p/k \rightarrow 0$ limit, $V(q,k)$ becomes independent of $p$ and Eq.~(\ref{eq:1loopmohammed}) can be directly compared to Eq.~(\ref{eq:Cov1loop}). One important difference is that in Eq.~(\ref{eq:Cov1loop}), while performing the angle part of the loop integral, one averages the product of two $\R_2$ responses, whereas in Eq.~(\ref{eq:1loopmohammed}), one has the product of two angle-averaged responses. The term $P_m(k_1)P_m(k_2)/\Plin(k_1)/\Plin(k_2)$ in Eq.~(\ref{eq:1loopmohammed}) is added as a correction term that improves the accuracy on nonlinear scales, in a way similar to the correction employed in Eq.~(\ref{eq:treemohammed}) for the standard tree level covariance. Finally, the Lorentzian damping term $\big(1+(p/p_\text{nl})^2\big)^{-2}$ in Eq.~(\ref{eq:1loopmohammed}), with $p_\text{nl}$ being a free parameter, is introduced to help suppress the excess of amplitude of \refeq{1loopmohammed} on nonlinear scales (compared to simulations, as we shall see below). In our response-based description, a similar damping is included without any adjustable parameters by employing the fully nonlinear responses (see Fig.~1 in Ref.~\cite{paper1}). Thus, one can interpret the rescaling by $(P_m/\Plin)$ and the phenomenological damping term in \refeq{1loopmohammed} as a rough model of the physical nonlinear response.

\begin{figure}
	\centering
	\includegraphics[width=\textwidth]{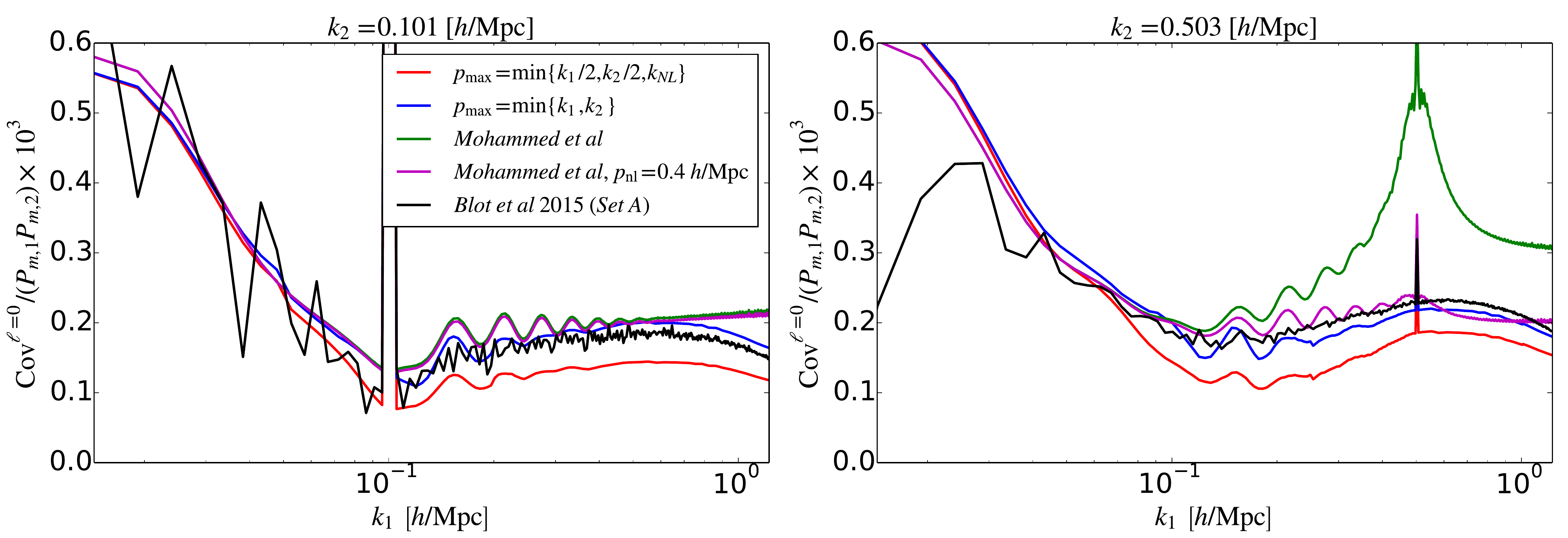}
	\caption{Comparison between our covariance prediction and that of the model of Ref.~\cite{mohammed1}. The two panels show the total $z=0$ angle-averaged covariance matrix as a function of $k_1$, for the values of $k_2$ indicated above each panel. Our results are shown for two $p_\text{max}$ settings (red and blue curves), as labeled. The one depicted by the red line is the theoretically self-consistent case that we used in the main body of the paper. The result from Ref.~\cite{mohammed1} (labeled as $Mohammed\ et\ al$) is shown with (magenta) and without (green) the phenomenological Lorentzian damping term in the 1-loop contribution.}
\label{fig:mohammedcomp}
\end{figure}

Figure \ref{fig:mohammedcomp} compares our predictions with those of Ref.~\cite{mohammed1}, labeled as $Mohammed\ et\ al$. The two panels show the total angle-averaged covariance as a function of $k_1$, for the fixed values of $k_2$ indicated above each panel. The red curve displays the result from our calculation as described in the main body of this paper. The model of Ref.~\cite{mohammed1} is shown for two choices of $p_\text{nl}$: $p_\text{nl} = \infty$ (green) and $p_\text{nl} = 0.4\kunit$ (magenta). In the left panel, for $k_2 = 0.1\kunit$, the model of Ref.~\cite{mohammed1} captures the amplitude of the simulation results better, even though the agreement is not perfect. This level of agreement with the simulation results for $k_1 \gtrsim 0.1\kunit$ depends sensitively on the correction factor $P_m(k_1)P_m(k_2)/\Plin(k_1)/\Plin(k_2)$ in Eqs.~(\ref{eq:treemohammed}) and (\ref{eq:1loopmohammed}). Concretely, dropping the correction term results in a much lower amplitude of the curves whenever $k_\text{hard} \gtrsim \knl$.

On the other hand,  the right panel ($k_2 \approx 0.5\kunit$) shows that the model of Ref.~\cite{mohammed1}, specifically the contribution in \refeq{1loopmohammed}, drastically overpredicts the simulation results on small scales (green) if the damping term is not included. This forces one to tune the value of the parameter $p_\text{nl}$ to bring the model closer to the simulations (magenta). As we have seen in the main body of the paper, our calculation (red line) still underpredicts the simulation results on nonlinear scales. However, the well-defined physical grounds of the response approach allows us to identify higher loops (which cannot be captured by a term of the type of \refeq{1loopmohammed}) as a likely reason for this discrepancy. 

As an exercise, we show as the blue curve the result of our calculation when allowing for a larger loop-momentum cutoff $p_\text{max}$. Specifically, the loop momentum is allowed to be as large as $k_\text{soft}$ and larger than $\knl$, at which point the loop integrand is strictly no longer in the response regime. This can be seen as a phenomenological attempt to capture higher-loop contributions that are currently left out. Although it is interesting to observe that the blue curve is capable to describe the simulation results quite well (for these values of $k_2$, at least), we stress that such a calculation is not guaranteed to be self-consistent, and that it should not be used to make predictions based on the response approach. Allowing $p_\text{max}$ to be a free fitting parameter risks \emph{overfitting} the simulation measurements, and loses the predictivity of the covariance for other cosmologies. These are issues that generally apply to any model that contains free fitting parameters.

\bibliography{REFS}

\end{document}